\documentclass[a4paper,12pt]{article}
\usepackage{amsmath, amssymb, latexsym, amscd, amsthm,amsfonts,amstext}
\usepackage[mathscr]{eucal}
\usepackage{graphicx}
\usepackage{subfig}

\numberwithin{equation}{section}
\input{tcilatex}
\begin{document}

\title{Stability Estimates for Some Parabolic Inverse Problems With the
Final Overdetermination Via a New Carleman Estimate}
\author{ Michael V. Klibanov \and $^{\ast }$Department of Mathematics and
Statistics, University of North \and Carolina at Charlotte, Charlotte, NC
28223, USA \and \texttt{mklibanv@charlotte.edu}}
\date{}
\maketitle

\begin{abstract}
This paper is about H\"{o}lder and Lipschitz stability estimates and
uniqueness theorems for some coefficient inverse problems and associated
inverse source problems for a general linear parabolic equation of the
second order with variable coefficients. The data for the inverse problem
are given at the final moment of time $\left\{ t=T\right\} $. In addition,
both Dirichlet and Neumann boundary conditions are given either on a part or
on the entire lateral boundary. Thus, if these boundary conditions are given
only at a part of the boundary, then even if the target coefficient is
known, still the forward problem is not a classical initial boundary value
problem.
\end{abstract}



\section{Introduction}

\label{sec:1}

It is well known that any Coefficient Inverse Problem (CIP) is nonlinear. On
the other hand, to prove uniqueness and stability theorems for a CIP, it is
convenient to consider a linear Inverse Source Problem (ISP), which is
directly generated by that CIP. We consider two CIPs and two corresponding
ISPs for a general parabolic Partial Differential Equation (PDE) of the
second order with $x-$dependent variable coefficients, where $x\in \Omega
\subset \mathbb{R}^{n}$ and $\Omega $ is a bounded domain. Our goal is to
prove either H\"{o}lder or Lipschitz stability estimates for our CIPs and
ISPs.

Let time $t\in \left( 0,T\right) .$ We assume in our CIP that we know the
initial condition at $\left\{ t=0\right\} $ and the lateral Cauchy data,
i.e. both Dirichlet and Neumann boundary conditions are known at a part $%
\Gamma \times \left( 0,T\right) $ of the lateral boundary $\partial \Omega
\times \left( 0,T\right) $, where $\Gamma \subseteq \partial \Omega .$
However, currently, a methodology, which would allow one to prove stability
results for such CIPs does not exist. And even uniqueness results are very
limited, see subsection 2.4 for some details. Furthermore, it is a long
standing open problem to obtain stability results for such a case. This the
reason why do we assume that, in addition to the above data, we know the
solution of that PDE at $\left\{ t=T\right\} ,$ i.e. we assume the final
overdetermination.

In the first CIP and, respectively, in the first ISP, we assume that both
Dirichlet and Neumann boundary conditions are given at any small part of the
lateral boundary of the domain of interest. Recall that a forward problem
for a PDE is a problem of finding a solution of this PDE in the case when
all coefficients of this equation are known. In the conventional case, a
forward problem for a parabolic PDE is one of classical initial boundary
value problems \cite{Ladpar}. However, the forward problem for our first
CIP/ISP is not a classical initial boundary value problem. In our second
CIP/ISP both Dirichlet and Neumann boundary conditions are given at the
entire lateral boundary $\partial \Omega \times \left( 0,T\right) $. Hence,
the forward problem in the second case is a classical initial boundary value
problem.

We prove the H\"{o}lder stability estimate and uniqueness theorem for our
first ISP. Next, we prove a stronger Lipschitz stability estimate for our
second ISP. Corresponding stability and uniqueness results for our CIPs are
not formulated here for brevity since they follow immediately from those for
ISPs.

We note that our input data are non overdetermined ones. Indeed, it is well
known that an input data are called non-overdetermined if the number $m$ of
free variables in the data equals the number $n$ of free variables in the
unknown coefficient, $m=n$. If, however, $m>n$, then such input data are
called overdetermined. In our case $m=n$.

A modified framework of \cite{BukhKlib} is used here. In \cite{BukhKlib},
Carleman estimates were introduced in the field of CIPs, see, e.g. \cite%
{ImYam1,ImYam2,Isakov,Klib84,Klib92,Ksurvey,KL,Lay} for some samples of
publications, which use the framework of \cite{BukhKlib}. The idea of \cite%
{BukhKlib} is also applicable to the numerical aspect for CIPs, which is
important for applications. More precisely, a globally convergent
convexification numerical method for CIPs was developed using a modified
idea of \cite{BukhKlib}, see, e.g. \cite{KLZ}, \cite[Chapter 9]{KL} for the
application of this method to the numerical solution of a CIP for a
parabolic PDE.

Our inverse problems have applications in the heat conduction theory \cite%
{Alifanov}. Another application is in the diffuse medical optics since
photons propagate in the diffuse manner in biological tissues \cite{Das}. In
the case of our inverse problems, one wants to figure out the history of the
process via measuring at the final moment of time either the spatial
distribution of the temperature or the spatial distribution of photons. In
addition, one is measuring either the temperature or the density of photons
at a part of the boundary as well as their fluxes at that part. If
coefficients of the corresponding elliptic operator are known, then this is
a well known problem of the solution of the parabolic equation in the
reversed direction in time. In this case one needs to know either only
Dirichlet or Neumann boundary condition at the entire lateral boundary. Even
though this is an unstable problem, there are some regularization methods
for it, see, e.g. \cite{KY}. We, however, consider the cases when either one
of the coefficients of that elliptic operator is unknown (CIP) or the source
function is unknown (ISP).

In section 2 we discuss new elements of this paper as well as known results
for CIPs and ISPs for parabolic equations, In section 3 we state our CIPs
and ISPs. We formulate our theorems in section 4. Sections 5-7 are devoted
to proofs of these theorems. Our Carleman estimate is proven Appendix, which
is section 8. All functions considered below are real valued ones.

\section{New Elements of This Paper and Known Results}

\label{sec:2}

\subsection{New elements}

\label{sec:2.1}

Our main new idea here is to arrange the mutual cancellation of parasitic
integrals over $\left\{ t=0\right\} $ and $\left\{ t=T\right\} $, which
conventionally occur in Carleman estimates for parabolic operators, see,
e.g. \cite{Ksurvey}, \cite[theorems 2.3.1 and 9.4.1]{KL}, \cite[Chapter 4, 
\S 1]{LRS}. The only case when those parasitic integrals do not occur is the
Carleman estimate of \cite{ImYam1}, which, however, does not work for our
goal since the Carleman Weight Function (CWF) of \cite{ImYam1} decays
exponentially at $t\rightarrow 0^{+}$ and $t\rightarrow T^{-}.$

To arrange that cancellation, we prove a new Carleman estimate for the
parabolic operator, in which the CWF depends on $x$ and is independent on $t$%
. To the best knowledge of the author, CWFs in all known Carleman estimates
for parabolic operators depend on both $x$ and $t$, see, e.g. \cite%
{ImYam1,ImYam2,Ksurvey}, \cite[section 2.3]{KL} and \cite[Chapter 4, \S 1]%
{LRS}.

\textbf{Remarks 2.1:}

\begin{enumerate}
\item \emph{We point out that the main goal of this paper is to present the
above new idea of the mutual cancellation of those parasitic terms. Thus,
although our results can be generalized in a number of different ways, we
are not concerned with such generalizations here. }

\item \emph{For the same reason we are not concerned here with extra
smoothness conditions. In particular, we recall that extra smoothness is a
minor issue in studies of CIPs, see, e.g. \cite{Nov1,Nov2}, \cite[Theorem 4.1%
]{Rom}. }
\end{enumerate}

\subsection{Known results for CIPs for parabolic equations with the final
overdetermination}

\label{sec:2.2}

Prior to the current work, CIPs for parabolic equations with the final
overdetermination were considered in \cite{ImYam2,Isakov,Prilepko}.
Lipschitz stability estimates for the CIP for the parabolic PDE with the
final overdetermination at $\left\{ t=T\right\} $ were obtained by Isakov 
\cite[section 9.1]{Isakov} and Prilepko, Orlovsky and Vasin \cite[section 1.2%
]{Prilepko}. In both these references the Dirichlet boundary condition is
given on the entire lateral boundary, and the Neumann boundary condition is
not given on any part of that boundary. It is assumed in \cite%
{Isakov,Prilepko} that the Dirichlet boundary value problem for the
associated elliptic operator has at most one solution.

In the recent publication of Imanuvilov and Yamamoto \cite{ImYam2}, the
Lipschitz stability estimate for a CIP for a parabolic PDE with the final
overdetermination at $\left\{ t=T\right\} $ is obtained. This estimate is
proven in \cite{ImYam2} only for the coefficient $c\left( x\right) $ in the
term $c\left( x\right) u$ of the elliptic operator, see section 3 for this
operator. The zero Neumann boundary condition is given at the entire
boundary in \cite{ImYam2}. In addition, the Dirichlet boundary condition is
given at any small part of the boundary. Convergence arguments are used in 
\cite{ImYam2} to prove the Lipschitz stability estimate.

\subsection{The main differences between our paper and \protect\cite%
{ImYam2,Isakov,Prilepko}}

\label{sec:2.3}

The first main difference is our above mentioned idea (subsection 2.1) of
arranging the mutual cancellation of parasitic integrals over $\left\{
t=0\right\} $ and $\left\{ t=T\right\} .$ This idea was not used in \cite%
{ImYam2,Isakov,Prilepko}.

In turn, this idea of ours causes the second important difference of our
case with \cite{ImYam2,Isakov,Prilepko}: we use a new Carleman estimate, in
which the CWF is independent on $t$.

The third main difference of our paper with the works \cite%
{ImYam2,Isakov,Prilepko} is our H\"{o}lder stability estimate. Indeed,
unlike these publications, in the H\"{o}lder stability estimate of this
paper, both the Dirichlet and Neumann boundary conditions are given only at
any small part of the lateral boundary.

The fourth main difference is the one with \cite{Isakov,Prilepko}. More
precisely, unlike these references, we do not impose the assumption
mentioned in subsection 2.2 that the Dirichlet boundary value problem for
the associated elliptic operator has at most one solution.

In fifth main difference is the one with \cite{ImYam2}. Indeed, we obtain
both H\"{o}lder and Lipschitz stability estimates for any coefficient of the
elliptic operator, which is unlike \cite{ImYam2}, where the Lipschitz
stability estimate is obtained only for the coefficient $c\left( x\right) $
in the zero order term $c\left( x\right) u\left( x,t\right) $ is estimated.

Finally, the sixth main difference is that, unlike \cite{ImYam2}, we do not
use convergence arguments in our proofs. In fact, these arguments are not
desirable ones for possible future numerical studies of our CIPs via the
so-called \textquotedblleft convexification method", as in, e.g. \cite{KLZ}, 
\cite[Chapter 9]{KL}.

\subsection{Three types of other known stability and uniqueness results for
CIPs and ISPs for parabolic PDEs}

\label{sec:2.4}

We now list three types of stability and uniqueness results for the CIPs for
parabolic PDEs, which are known so far, in addition to the above cited ones 
\cite{ImYam2,Isakov,Prilepko}. All results listed below are obtained using
the framework of \cite{BukhKlib}. We refer to \cite[section 3.4]{KL} for a
detailed discussion of the topic of this subsection.

First, this is the case when the solution of the parabolic equation is known
at $\left\{ t=t_{0}\right\} ,$ where $0<t_{0}<T$ \cite{Ksurvey}, \cite[%
Theorem 3.4.3]{KL}. The Dirichlet and Neumann boundary conditions on a part
of the lateral boundary are also known in this case, and the initial
condition at $\left\{ t=0\right\} $ is unknown. Even though only uniqueness
theorems were obtained in \cite{Ksurvey,KL}, H\"{o}lder stability estimates
can be obtained as well via small modifications of those proofs. In \cite%
{ImYam1} a stronger Lipschitz stability estimate was obtained for this
problem. 

Second, this is the case when the forward problem for the parabolic equation
is the Cauchy problem in $\mathbb{R}^{n}\times \left( 0,T\right) $, the
solution of this forward problem is known at $\left\{ t=T\right\} ,$ and the
target coefficient is known in an arbitrary domain $\omega \subset \mathbb{R}%
^{n}$ and unknown in $\mathbb{R}^{n}\diagdown \omega $ \cite[Theorem 1]%
{BukhKlib}, \cite{Klib84,Klib92,Ksurvey}, \cite[Theorem 3.4.4]{KL}. Again,
the initial condition at $\left\{ t=0\right\} $ is unknown in these
publications. In this case the so-called Reznickaya transform \cite[formula
(3.100)]{KL}, \cite[formula (7.129)]{LRS} can be used to prove the
analyticity of the solution of that forward problem as the function of the
real variable $t>0$. Then the knowledge in $\omega $ of the target
coefficient combined with that analyticity leads to the knowledge of both
Dirichlet and Neumann boundary conditions at $\partial \omega \times \left(
0,\infty \right) .$ Next, results of the above first case are applicable to
obtain uniqueness theorems.

The Reznickaya transform is one-to-one. This is a modified Laplace
transform, which transforms the solution of the Cauchy problem for the
hyperbolic equation in the solution of the Cauchy problem for the similar
parabolic equation.

Third, this is the case when the initial condition at $\left\{ t=0\right\} $
is known and the forward problem is the Cauchy problem for the parabolic
equation \cite{Klib92,Ksurvey}, \cite[Theorem 3.4.2]{KL}. The main fact,
which is used in these works, is that the original CIP is connected with the
CIP for the analogous hyperbolic equation via the above mentioned Reznickaya
transform. Since this transform is one-to-one, then the uniqueness theorem
for the original CIP for the parabolic equation follows from the uniqueness
theorem for the corresponding CIP for that hyperbolic equation. On the other
hand, since the inverse Reznickaya transform is a modified inverse Laplace
transform and since the latter is very unstable, then valuable stability
results for the parabolic case cannot be obtained this way.

\section{Statements of the Coefficient Inverse Problems and the Inverse
Source Problems}

\label{sec:3}

We denote $x=\left( x_{1},x_{2},...,x_{n}\right) =\left( x_{1},\overline{x}%
\right) $ points in $\mathbb{R}^{n}.$ Let $\Omega \subset \mathbb{R}^{n}$ be
a bounded domain with a piecewise smooth boundary with $C^{6}-$pieces. Let
the number $T>0$ and let $\Gamma \subseteq \partial \Omega ,\Gamma \in C^{6}$
be a part of the boundary of the domain $\Omega .$ Denote 
\begin{equation}
Q_{T}=\Omega \times \left( 0,T\right) ,S_{T}=\partial \Omega \times \left(
0,T\right) ,\Gamma _{T}=\Gamma \times \left( 0,T\right) .  \label{1.07}
\end{equation}

Let functions%
\begin{equation}
a^{ij}\left( x\right) \in C^{1}\left( \overline{\Omega }\right) ,\text{ }%
i,j=1,...n,  \label{1.0}
\end{equation}%
\begin{equation}
a^{ij}\left( x\right) =a^{ji}\left( x\right) ,\text{ }i,j=1,...n,
\label{1.3}
\end{equation}%
\begin{equation}
A=\max_{i,j}\left\Vert a_{ij}\right\Vert _{C^{1}\left( \overline{\Omega }%
\right) },  \label{3.140}
\end{equation}%
\begin{equation}
\nu \left\vert \eta \right\vert ^{2}\leq
\dsum\limits_{i,j=1}^{n}a^{ij}\left( x\right) \eta _{i}\eta _{j},\text{ }%
\forall x\in \overline{\Omega },\forall \eta \in \mathbb{R}^{n},  \label{1.4}
\end{equation}%
\begin{equation}
b_{j}\left( x\right) ,c\left( x\right) \in C\left( \overline{\Omega }\right)
,\text{ }j=1,...,n,  \label{1.1}
\end{equation}%
where $\nu >0$ is a number. For any appropriate function $u\left( x,t\right) 
$ denote%
\begin{equation}
L_{0}u=\dsum\limits_{i,j=1}^{n}a^{ij}\left( x\right) u_{x_{i}x_{j}},
\label{1.5}
\end{equation}%
\begin{equation}
Lu=\dsum\limits_{i,j=1}^{n}a^{ij}\left( x\right)
u_{x_{i}x_{j}}+\dsum\limits_{j=1}^{n}b_{j}\left( x\right) u_{x_{j}}+c\left(
x\right) u=L_{0}u+L_{1}u.  \label{1.6}
\end{equation}%
Let the function $u\in C^{6,3}\left( \overline{Q}_{T}\right) $ satisfies the
following conditions:%
\begin{equation}
u_{t}=Lu\text{ in }Q_{T},  \label{1.7}
\end{equation}%
\begin{equation}
u\left( x,0\right) =f\left( x\right) \text{ in }\Omega ,  \label{1.8}
\end{equation}%
\begin{equation}
u\mid _{\Gamma _{T}}=p\left( x,t\right) ,  \label{1.10}
\end{equation}%
\begin{equation}
\partial _{n}u\mid _{\Gamma _{T}}=q\left( x,t\right) ,  \label{1.01}
\end{equation}%
\begin{equation}
u\left( x,T\right) =F\left( x\right) \text{ in }\Omega ,  \label{1.9}
\end{equation}%
where $n$ in the outward unit normal vector on $\Gamma $. Hence, we must
have functions $f,F\in C^{6}\left( \overline{\Omega }\right) .$ Also, $%
\partial _{t}-L_{0}$ is the principal part of the parabolic operator $%
\partial _{t}-L.$ When formulating inverse problems, we differentiate two
cases:%
\begin{equation}
\Gamma \neq \partial \Omega ,\text{ incomplete boundary data,}  \label{1.90}
\end{equation}%
\begin{equation}
\Gamma =\partial \Omega ,\text{ complete boundary data.}  \label{1.91}
\end{equation}
Note that since, in general at least, $\Gamma \neq \partial \Omega $, then
neither problem (\ref{1.7})-(\ref{1.10}) nor problem (\ref{1.7}), (\ref{1.8}%
), (\ref{1.01}) is not necessary the initial boundary value problem for
equation (\ref{1.7}). We consider the following inverse problem:

\textbf{Coefficient Inverse Problem 1 (CIP1, incomplete boundary data)}\emph{%
.\ Assume that one of coefficients of the operator }$L$\emph{\ in (\ref{1.6}%
) is unknown and all other coefficients are known. Suppose that (\ref{1.90})
holds. Determine that unknown coefficient for }$x\in \Omega ,$\emph{\
assuming that functions }$f\left( x\right) ,F\left( x\right) ,p\left(
x,t\right) $\emph{\ and }$q\left( x,t\right) $\emph{\ in (\ref{1.8})-(\ref%
{1.9}) are known. }

\textbf{Coefficient Inverse Problem 2 (CIP2, complete boundary data)}\emph{.
Suppose that conditions of\ CIP1 hold, except that (\ref{1.90}) is replaced
with (\ref{1.91}). Determine that unknown coefficient for }$x\in \Omega ,$%
\emph{\ assuming that functions }$f\left( x\right) ,F\left( x\right)
,p\left( x,t\right) $\emph{\ and }$q\left( x,t\right) $\emph{\ in (\ref{1.8}%
)-(\ref{1.9}) are known. }

Due to (\ref{1.9}) these are the CIPs with the final overdetermination. It
is well known that in order to prove stability and uniqueness results for
either of our two CIPs, it is sufficient to prove such results for
associated ISPs. And then the corresponding results for CIPs follow
immediately. To derive these ISPs from our CIPs, we proceed via the well
known way. For example, let the coefficient $a^{i_{0}j_{0}}\left( x\right) $
be unknown in either of our CIPs. It is well known that in order to get a
stability estimate for this problem, we need to consider two pairs of
functions $\left( u_{1}\left( x,t\right) ,a_{1}^{i_{0}j_{0}}\left( x\right)
\right) $ and $\left( u_{2}\left( x,t\right) ,a_{2}^{i_{0}j_{0}}\left(
x\right) \right) .$ Keeping in mind that by (\ref{1.3}) $a^{i_{0}j_{0}}%
\left( x\right) =a^{j_{0}i_{0}}\left( x\right) $ and assuming that $%
i_{0}\neq j_{0},$ denote 
\begin{equation*}
\widetilde{u}\left( x,t\right) =u_{1}\left( x,t\right) -u_{2}\left(
x,t\right) ,\text{ }b\left( x\right) =a_{1}^{i_{0}j_{0}}\left( x\right)
-a_{2}^{i_{0}j_{0}}\left( x\right) .
\end{equation*}%
Let $L^{\left( 1\right) }$ be the operator $L$ in (\ref{1.6}) in the case
when the coefficients $a^{i_{0}j_{0}}\left( x\right) =$ $a^{j_{0}i_{0}}%
\left( x\right) $ are replaced with $a_{1}^{i_{0}j_{0}}\left( x\right)
=a_{1}^{j_{0}i_{0}}\left( x\right) .$ Then equation (\ref{1.7}) implies%
\begin{equation}
\widetilde{u}_{t}-L^{\left( 1\right) }\widetilde{u}=b\left( x\right) \left(
-2u_{2x_{i_{0}}x_{j_{0}}}\right) \text{ in }Q_{T}.  \label{1.11}
\end{equation}%
If $i_{0}=j_{0},$ then the multiplier \textquotedblleft 2" should not be
present in (\ref{1.11}). The case when any coefficient of the operator $%
L_{1} $ in the lower order terms of the (\ref{1.6}) is completely similar.
Hence, it is convenient to introduce the function $R\left( x,t\right) $ and
to consider the following inverse source problems (slightly abusing the
above notations):

\textbf{Inverse Source Problem 1 (ISP1, incomplete boundary data).} \emph{%
Assume that condition (\ref{1.90}) holds. Let the function }$R\left(
x,t\right) \in C^{6,3}\left( \overline{Q}_{T}\right) $\emph{\ and the
function }$b\left( x\right) \in C^{1}\left( \overline{\Omega }\right) .$%
\emph{\ Let the function }$u\left( x,t\right) \in C^{6,3}\left( \overline{Q}%
_{T}\right) $\emph{\ satisfies the following conditions:}%
\begin{equation}
u_{t}=Lu+b\left( x\right) R\left( x,t\right) \text{ in }Q_{T},  \label{1.12}
\end{equation}%
\begin{equation}
u\left( x,0\right) =f\left( x\right) \text{ in }\Omega ,  \label{1.13}
\end{equation}%
\begin{equation}
u\left( x,T\right) =F\left( x\right) \text{ in }\Omega ,  \label{1.14}
\end{equation}%
\begin{equation}
u\mid _{\Gamma _{T}}=p\left( x,t\right) ,\partial _{n}u\mid _{\Gamma
_{T}}=q\left( x,t\right) .  \label{1.15}
\end{equation}%
\emph{Suppose that all coefficients of the operator }$L$\emph{, the function 
}$R$\emph{\ and the right hand sides of (\ref{1.13})-(\ref{1.15}) are known,
but the function }$b\left( x\right) $\emph{\ is unknown. Estimate the
function }$b\left( x\right) $\emph{\ via functions involved in the right
hand sides of (\ref{1.13})-(\ref{1.15}).}

\textbf{Inverse Source Problem 2 (ISP2, complete boundary data). }\emph{%
Assume that condition (\ref{1.91}) holds and the rest of conditions of ISP1
are valid. Estimate the function }$b\left( x\right) $\emph{\ via functions
involved in the right hand sides of (\ref{1.13})-(\ref{1.15}).}

Hence, ISP1 is generated by CIP1 and ISP2 is generated by CIP2. Note that by
(\ref{1.11})%
\begin{equation*}
R\left( x,t\right) =\left\{ 
\begin{array}{c}
-2u_{2x_{i_{0}}x_{j_{0}}}\text{ if }i_{0}\neq j_{0}, \\ 
-u_{2x_{i_{0}}x_{i_{0}}}\text{ if }i_{0}=j_{0}.%
\end{array}%
\right.
\end{equation*}%
We assume below that 
\begin{equation}
\left\vert R\left( x,t\right) \right\vert \geq \sigma \text{ in }\overline{Q}%
_{T},  \label{1.16}
\end{equation}%
where $\sigma >0$ is a number. We now briefly discuss some sufficient
conditions, in terms of the above CIPs, which ensure (\ref{1.16})\emph{. }In
the above example, which led to (\ref{1.11}), it is sufficient to assume
that the initial condition $f\left( x\right) $ in (\ref{1.8}) is such that $%
f\in C^{6}\left( \overline{\Omega }\right) ,f_{x_{i_{0}}x_{j_{0}}}\left(
x\right) \neq 0$ in $\overline{\Omega }$ and $T$ is sufficiently small. The
second scenario ensuring (\ref{1.16}) is for CIP2. In this case one needs to
assume that condition (\ref{1.91}) holds, implying $\Gamma _{T}=S_{T},$ in (%
\ref{1.6}) the coefficient $c\left( x\right) $ is unknown, $c\left( x\right)
\leq 0$ , $f\left( x\right) \geq \sigma $ in $\overline{\Omega },$ and the
Dirichlet boundary condition $p\left( x,t\right) $ in (\ref{1.10}) is \ such
that $p\left( x,t\right) \geq \sigma $ on $S_{T}.$In this case, (\ref{1.16})
follows from the maximum principle. As to the required smoothness of
functions $u\left( x,t\right) ,R\left( x,t\right) \in C^{6,3}\left( 
\overline{Q}_{T}\right) ,$ we refer to Remarks 2.1.

\section{Theorems}

\label{sec:4}

To reduce the number of notations, we introduce below numbers rather than
symbols when specifying the geometrical parameters characterizing the domain 
$\Omega .$ Without any loss of the generality we assume that 
\begin{equation}
\Gamma =\left\{ x_{1}=0,\left\vert \overline{x}\right\vert <1\right\}
\subset \partial \Omega \text{ in the case (\ref{1.90}).}  \label{3.1}
\end{equation}%
Indeed, we can always assume that there exists a piece $\Gamma ^{\prime
}\subseteq \Gamma ,$ which can be parametrized as 
\begin{equation}
\Gamma ^{\prime }=\left\{ x_{1}=s\left( \overline{x}\right) ,\left\vert 
\overline{x}\right\vert <\theta \right\} ,s\left( \overline{x}\right) \in
C^{6}\left( \left\vert \overline{x}\right\vert \leq \theta \right) ,
\label{3.01}
\end{equation}%
where the positive number $\theta $ is sufficiently small. Changing
variables 
\begin{equation}
\left( x_{1},\overline{x}\right) \Leftrightarrow \left( x_{1}^{\prime },%
\overline{x}^{\prime }\right) =\left( x_{1}^{\prime }=x_{1}-s\left( 
\overline{x}\right) ,\text{ }\overline{x}^{\prime }=\frac{\overline{x}}{%
\theta }\right)   \label{3.100}
\end{equation}%
and keeping the same notations for brevity, we obtain (\ref{3.1}). Thus, by (%
\ref{3.1}), we assume that the domain $G\subset \Omega ,$%
\begin{equation}
G=\left\{ x_{1}+\frac{\left\vert \overline{x}\right\vert ^{2}}{2}+\frac{1}{4}%
<\frac{3}{4},\text{ }x_{1}>0\right\} \subset \Omega .  \label{3.2}
\end{equation}%
Thus, by (\ref{3.1}) and (\ref{3.2}) 
\begin{equation}
\Gamma \subset \partial G\text{ in the case (\ref{1.90}).}  \label{3.200}
\end{equation}

\textbf{Remark 4.1}. \emph{Thus, it follows from (\ref{1.15}) and (\ref{3.1}%
)-(\ref{3.200}) that the lateral Cauchy data in Theorem 2 (below) can be
given at any small part of the boundary of the domain }$\Omega .$

Let $\mu \geq 1$ and $\lambda \geq 1$ be two parameters, which we define
later. Introduce two functions $\varphi \left( x\right) $ and $\phi \left(
x\right) ,$%
\begin{equation}
\varphi \left( x\right) =x_{1}+\frac{\left\vert \overline{x}\right\vert ^{2}%
}{2}+\frac{1}{4},\text{ }x\in G,  \label{3.3}
\end{equation}%
\begin{equation}
\phi \left( x\right) =e^{\lambda \varphi ^{-\mu }}.  \label{3.4}
\end{equation}%
Hence, by (\ref{3.2}) and (\ref{3.3}) 
\begin{equation}
\left\{ \frac{1}{4}<\varphi \left( x\right) <\frac{3}{4},x_{1}>0\right\} =G.
\label{3.5}
\end{equation}%
Similar functions $\varphi $ and $\phi $ are used in conventional Carleman
estimates for parabolic operators, see, e.g. \cite[section 2.3]{KL}, \cite[%
Chapter 4, \S 1]{LRS}. However, if following \cite{KL,LRS}, then the $t-$%
dependent term $\left( t-T/2\right) ^{2}/\left( 2T^{2}\right) $ should be
added to the function $\varphi \left( x\right) .$ Also, these functions $%
\varphi $ and $\phi $ are used in Carleman estimates for elliptic operators 
\cite[section 2.4]{KL}, \cite[Chapter 4, \S 1]{LRS}. Choose a number $%
\varepsilon ,$ 
\begin{equation}
\varepsilon \in \left( 0,1/2\right) .  \label{3.50}
\end{equation}%
Denote%
\begin{equation}
\left. 
\begin{array}{c}
G_{\varepsilon }=\left\{ \varphi \left( x\right) <3/4-\varepsilon ,\text{ }%
x_{1}>0\right\} = \\ 
=\left\{ x_{1}+\left\vert \overline{x}\right\vert ^{2}/2+1/4<3/4-\varepsilon
,\text{ }x_{1}>0\right\} \subset G.%
\end{array}%
\right.  \label{3.06}
\end{equation}%
Denote 
\begin{equation}
G_{T}=G\times \left( 0,T\right) ,  \label{3.6}
\end{equation}%
\begin{equation}
G_{\varepsilon ,T}=G_{\varepsilon }\times \left( 0,T\right) ,  \label{3.60}
\end{equation}%
\begin{equation}
\partial _{1}G=\Gamma =\left\{ \varphi \left( x\right) <\frac{3}{4}%
,x_{1}=0\right\} ,\text{ }\partial _{1}G_{T}=\Gamma _{T}=\Gamma \times
\left( 0,T\right) ,  \label{3.7}
\end{equation}%
\begin{equation}
\partial _{2}G=\left\{ \varphi \left( x\right) =\frac{3}{4},x_{1}>0\right\} ,%
\text{ }\partial _{2}G_{T}=\partial _{2}G\times \left( 0,T\right) ,
\label{3.8}
\end{equation}%
\begin{equation}
\partial _{1}G_{\varepsilon }=\Gamma _{\varepsilon }=\left\{ \varphi \left(
x\right) <\frac{3}{4}-\varepsilon ,x_{1}=0\right\} ,\text{ }\partial
_{1}G_{\varepsilon ,T}=\Gamma _{\varepsilon ,T}=\Gamma _{\varepsilon }\times
\left( 0,T\right) ,  \label{3.9}
\end{equation}%
\begin{equation}
\partial _{2}G_{\varepsilon }=\left\{ \varphi \left( x\right) =\frac{3}{4}%
-\varepsilon \right\} ,\text{ }\partial _{2}G_{\varepsilon ,T}=\partial
_{2}G_{\varepsilon }\times \left( 0,T\right) .  \label{3.10}
\end{equation}%
Hence, by (\ref{3.2}) and (\ref{3.06})-(\ref{3.10}) 
\begin{equation}
\partial G=\Gamma \cup \partial _{2}G,  \label{3.11}
\end{equation}%
\begin{equation}
\partial G_{\varepsilon }=\Gamma _{\varepsilon }\cup \partial
_{2}G_{\varepsilon }.  \label{3.110}
\end{equation}%
By (\ref{3.3})-(\ref{3.5}) 
\begin{equation}
\max_{\overline{G}}\phi ^{2}\left( x\right) =\phi ^{2}\left( 0\right) =\exp
\left( 2\lambda \left( \frac{1}{4}\right) ^{-\mu }\right) ,  \label{3.111}
\end{equation}%
\begin{equation}
\min_{\overline{G}_{\varepsilon }}\phi ^{2}\left( x\right) =\phi ^{2}\left(
x\right) \mid _{\partial _{2}G_{\varepsilon }}=\exp \left( 2\lambda \left( 
\frac{3}{4}-\varepsilon \right) ^{-\mu }\right) >\exp \left( 2\lambda \left( 
\frac{3}{4}\right) ^{-\mu }\right) .  \label{3.12}
\end{equation}%
Furthermore, 
\begin{equation}
\phi ^{2}\left( x\right) =\exp \left( 2\lambda \left( \frac{3}{4}\right)
^{-\mu }\right) \text{ for }x\in \partial _{2}G_{T}.  \label{3.13}
\end{equation}

By (\ref{1.15}) and (\ref{3.7}) 
\begin{equation}
u\mid _{\Gamma _{T}}=p\left( x,t\right) ,\text{ }\partial _{n}u\mid _{\Gamma
_{T}}=q\left( x,t\right) .  \label{3.14}
\end{equation}
For brevity we denote below for any appropriate function $g\left( x,t\right)
:$ 
\begin{equation*}
g_{i}=\partial _{x_{i}}g,\text{ }\nabla g=\left( g_{1},...,g_{n}\right) ,%
\text{ }g_{ij}=\partial _{x_{i}x_{j}}^{2}g.
\end{equation*}

\subsection{The new Carleman estimate}

\label{sec:4.1}

Recall that the number $\nu >0$ is defined in (\ref{1.4}), and the operator $%
L_{0}$ is defined in (\ref{1.5}).

\textbf{Theorem 1} (pointwise Carleman estimate for the operator $\partial
_{t}-L_{0}).$ \emph{Assume that conditions (\ref{1.07})-(\ref{1.4}), (\ref%
{1.5}) and (\ref{3.1})-(\ref{3.4}) hold. Then there exist sufficiently large
numbers }$\mu _{0}=\mu _{0}\left( G,\nu ,A\right) \geq 1$\emph{\ and }$%
\lambda _{0}=\lambda _{0}\left( G,\nu ,A\right) \geq 1$\emph{\ as well as a
number }$C=C\left( G,\nu ,A\right) >0$\emph{\ depending only on listed
parameters such that the following pointwise Carleman estimate holds for }$%
\mu =\mu _{0},$ \emph{for all} $\lambda \geq \lambda _{0}$\emph{\ and for
all functions }$u\in C^{2,1}\left( \overline{G}_{T}\right) :$%
\begin{equation*}
\left( u_{t}-L_{0}u\right) ^{2}\phi ^{2}\geq \frac{C}{\lambda }\left(
u_{t}^{2}+\dsum\limits_{i,j=1}^{n}u_{i,j}^{2}\right) \phi ^{2}+C\lambda
\left( \nabla u\right) ^{2}\phi ^{2}+C\lambda ^{3}u^{2}\phi ^{2}+
\end{equation*}%
\begin{equation}
+V_{t}+\func{div}U,\text{ }\left( x,t\right) \in G_{T},  \label{3.15}
\end{equation}%
\emph{where the function }$V$\emph{\ is:}%
\begin{equation}
\left. 
\begin{array}{c}
\partial _{t}V= \\ 
==\partial _{t}\left( \left( d/2\right)
\dsum\limits_{i,j=1}^{n}a^{ij}\varphi ^{\mu _{0}+2}\left( u_{i}-\lambda \mu
_{0}\varphi _{i}\varphi ^{-\mu _{0}-1}u\right) \left( u_{j}-\lambda \mu
_{0}\varphi _{j}\varphi ^{-\mu _{0}-1}u\right) \phi ^{2}\right) + \\ 
+\partial _{t}\left( -\left( d/2\right) \lambda ^{2}\mu _{0}^{2}\varphi
^{-\mu _{0}}\dsum\limits_{i,j=1}^{n}a^{ij}\left( x\right) \left( \varphi
_{i}\varphi _{j}\left( 1-\lambda ^{-1}\left( 1+\mu _{0}^{-1}\right) \varphi
^{\mu }\right) \right) u^{2}\phi ^{2}\right) + \\ 
+\partial _{t}\left( \left( d/2\right) \lambda \mu _{0}u^{2}\phi ^{2}+\left(
4^{2\mu _{0}+2}\left( \lambda \mu _{0}\right) \right)
^{-1}\dsum\limits_{i,j=1}^{n}a^{ij}\left( x\right) u_{i}u_{j}\phi
^{2}\right) .%
\end{array}%
\right.  \label{3.16}
\end{equation}%
\emph{And }$\func{div}U$ \emph{of} \emph{the vector function }$U$\emph{\ is:}%
\begin{equation}
\left. 
\begin{array}{c}
\func{div}U=\left( d/2\right) \dsum\limits_{i,j=1}^{n}\left[ \left(
-a^{ij}w_{i}w_{t}\varphi ^{\mu _{0}+2}\right) _{j}+\left(
-a^{ij}w_{j}w_{t}\varphi ^{\mu _{0}+2}\right) _{i}\right] + \\ 
+\left( d/2\right) \dsum\limits_{i,j,k,s=1}^{n}\left[ 
\begin{array}{c}
\left( \lambda \mu _{0}a^{ij}\left( x\right) a^{ks}\left( x\right)
w_{i}w_{k}\right) _{j}+\left( \lambda \mu _{0}a^{ij}\left( x\right)
a^{ks}\left( x\right) w_{j}w_{k}\right) _{i}+ \\ 
+\left( -\lambda \mu _{0}a^{ij}\left( x\right) a^{ks}\left( x\right)
w_{i}w_{j}\right) _{k}%
\end{array}%
\right] + \\ 
+\left( d/2\right) \dsum\limits_{i,j,k,s=1}^{n}\left( \varphi ^{-2\mu
_{0}-1}a^{ij}\left( x\right) a^{ks}\left( x\right) \varphi _{k}\varphi
_{s}\varphi _{j}\left( \left( 1-\lambda ^{-1}\left( 1+\mu _{0}^{-1}\right)
\varphi ^{\mu _{0}}\right) \right) \phi ^{2}u^{2}\right) _{i}+ \\ 
+\left( d/2\right) \left( \lambda \mu _{0}\right)
^{-1}\dsum\limits_{i,j,k,s=1}^{n}\left( \left( a^{ij}\left( x\right)
a^{ks}\left( x\right) \varphi ^{-\mu _{0}}\varphi _{k}\varphi _{s}\varphi
_{j}\varphi _{ij}\right) \phi ^{2}u^{2}\right) _{i}+ \\ 
+\left( d/2\right) \dsum\limits_{i,j,k,s=1}^{n}\left( \varphi ^{-2\mu
_{0}-1}a^{ij}\left( x\right) a^{ks}\left( x\right) \varphi _{k}\varphi
_{s}\varphi _{i}\left( \left( 1-\lambda ^{-1}\left( 1+\mu _{0}^{-1}\right)
\varphi ^{\mu _{0}}\right) \right) \phi ^{2}u^{2}\right) _{j}+ \\ 
+\left( d/2\right) \left( \lambda \mu _{0}\right)
^{-1}\dsum\limits_{i,j,k,s=1}^{n}\left( \left( a^{ij}\left( x\right)
a^{ks}\left( x\right) \varphi ^{-\mu _{0}}\varphi _{k}\varphi _{s}\varphi
_{j}\varphi _{ij}\right) \phi ^{2}u^{2}\right) _{j}+ \\ 
+\dsum\limits_{i,j=1}^{n}\left( -da^{ij}\left( x\right) u_{i}u\phi
^{2}\right) _{j}+\left[ d\left( 4^{2\mu _{0}+2}\left( \lambda \mu
_{0}\right) \right) ^{-1}\dsum\limits_{i,j=1}^{n}\left( -2a^{ij}\left(
x\right) u_{t}u_{i}\phi ^{2}\right) \right] _{j}+ \\ 
+\left[ d\left( 4^{2\mu _{0}+2}\left( \lambda \mu _{0}\right) \right)
^{-1}\dsum\limits_{i,j,k,s=1}^{n}\left( a^{ij}\left( x\right) a^{ks}\left(
x\right) u_{i}u_{ks}\phi ^{2}\right) \right] _{j}+ \\ 
+\left[ d\left( 4^{2\mu _{0}+2}\left( \lambda \mu _{0}\right) \right)
^{-1}\dsum\limits_{i,j,k,s=1}^{n}\left( -a^{ij}\left( x\right) a^{ks}\left(
x\right) u_{i}u_{sj}\phi ^{2}\right) \right] _{k},%
\end{array}%
\right.  \label{3.160}
\end{equation}

\emph{where }$w_{i}=\left( u\phi \right) _{i},w_{t}=u_{t}\phi $ \emph{and
the number }$d$\emph{\ is}%
\begin{equation}
d=\frac{1}{1+\left( 2\lambda \mu _{0}4^{2\mu _{0}+2}\right) }.  \label{3.161}
\end{equation}

For the convenience of the reader, we prove Theorem 1 in Appendix. We assume
in all other derivations below that this theorem holds true.

\textbf{Corollary.}\emph{\ The following implications hold:}%
\begin{equation*}
u\left( x,0\right) =u\left( x,T\right) \rightarrow V\left( x,0\right)
=V\left( x,T\right) \rightarrow \dint\limits_{G_{T}}\partial _{t}V\left(
x,t\right) dxdt=0.
\end{equation*}

\textbf{Proof of Corollary.} By (\ref{3.3}) and (\ref{3.4}) functions $%
\varphi \left( x\right) $ and $\phi \left( x\right) $ are independent on $t.$
Hence, if $u\left( x,0\right) =u\left( x,T\right) ,$ then (\ref{3.16})
implies that $V\left( x,0\right) =V\left( x,T\right) .$ $\square $

Using this Corollary, we arrange below in proofs of Theorems 2 and 4 the
mutual cancellation of parasitic integrals over $\left\{ t=0\right\} $ and $%
\left\{ t=T\right\} $ as stated in subsection 2.1.

\subsection{Comments about Theorem 1}

\label{sec:4.2}

Even though there are theorems, which are similar with Theorem 1, see, e.g. 
\cite[Theorem 2.3.1]{KL}, \cite[Chapter 4, \S 1]{LRS}, and their proofs are
similar with ours in Appendix, still the CWFs in all previous works depend
on both $x$ and $t$. On the other hand, in order to make Corollary work, we
need the independence of the CWF on $t$. Therefore, we have no choice but to
prove this theorem.

It is well known that there are two methods of proofs of Carleman estimates.
The first method uses symbols of operators, see, e.g. \cite[Theorem 3.2.1]%
{Isakov}. In the second method, pointwise Carleman estimates are derived,
see, e.g. \cite[Theorem 2.3.1]{KL}, \cite[Chapter 4, \S 1]{LRS}. The first
method provides rather simple and short proofs of Carleman estimates.
However, this method works only with zero boundary conditions, which in the
parabolic case include zero initial and terminal conditions at $\left\{
t=0\right\} $ and $\left\{ t=T\right\} $ respectively. This means that
formulas (\ref{3.16}) and (\ref{3.160}) cannot be derived from the first
method. On the other hand, formula (\ref{3.16}) is important for Corollary,
which, in turn is the key for our mutual cancellation idea. Next, the
importance of formula (\ref{3.160}) is that it enables us to work with
non-zero boundary conditions (\ref{1.15}) in proofs of Theorems 2 and 4.
These are the reasons why we use the second method.

It is well known that derivations of pointwise Carleman estimates, like,
e.g. the one of (\ref{3.15}), are inevitably space consuming. However, this
is the price one pays for the ability to work with the non-zero boundary
conditions. In our case, the latter means working with formulas (\ref{3.16})
and (\ref{3.160}). Therefore, we work in Appendix with this space consuming
derivation. 

Even though $u\in C^{2,1}\left( \overline{G}_{T}\right) $\ in Theorem 1, it
follows from the density arguments \qquad and trace theorem that integrating
(\ref{3.15}) (\ref{3.160}) over the domain $G_{T}$ and using Gauss formula
as well as (\ref{3.16}) and (\ref{3.160}),\ we obtain that the resulting
estimate is valid for all functions $u\in H^{3}\left( G_{T}\right) ,$ and
this is what we actually use in Theorems 2 and 4$.$

\subsection{Formulations of Theorems 2-4}

\label{sec:4.3}

To work with Theorem 2, we impose conditions, which are slightly more
general than the ones in (\ref{1.12})-(\ref{1.16}). More precisely, we
assume that analogs of (\ref{1.12})-(\ref{1.16}) are valid in the domain $%
G_{T}\subset Q_{T}$ \ rather than in the domain $Q_{T},$ 
\begin{equation}
u_{t}=Lu+b\left( x\right) R\left( x,t\right) \text{ in }G_{T},  \label{3.151}
\end{equation}%
\begin{equation}
u\left( x,0\right) =f\left( x\right) \text{ in }G,  \label{3.152}
\end{equation}%
\begin{equation}
u\left( x,T\right) =F\left( x\right) \text{ in }G,  \label{3.153}
\end{equation}%
\begin{equation}
u\mid _{\partial _{1}G_{T}}=p\left( x,t\right) ,\partial _{n}u\mid
_{_{\partial _{1}G_{T}}}=q\left( x,t\right) ,  \label{3.154}
\end{equation}%
\begin{equation}
\left\vert R\left( x,t\right) \right\vert \geq \sigma \text{ in }\overline{G}%
_{T},  \label{3.155}
\end{equation}%
see (\ref{3.1})-(\ref{3.3}) and (\ref{3.11}) for (\ref{3.154}).

\textbf{Theorem 2 }(H\"{o}lder stability estimate)\textbf{.} \emph{Assume
that conditions (\ref{1.07})-(\ref{1.6}) hold, in which the domain }$\Omega $
\emph{is replaced with the domain }$G$\emph{. Also, let conditions (\ref{3.1}%
)-(\ref{3.200}) hold. Let in (\ref{3.152})-(\ref{3.154}) }%
\begin{equation}
\left\Vert p_{t}\right\Vert _{H^{2,0}\left( \Gamma _{T}\right) },\left\Vert
q_{t}\right\Vert _{H^{1,0}\left( \Gamma _{T}\right) }\leq \delta ,
\label{3.18}
\end{equation}%
\begin{equation}
\left\Vert f\right\Vert _{H^{4}\left( G\right) },\left\Vert F\right\Vert
_{H^{4}\left( G\right) }\leq \delta ,  \label{3.19}
\end{equation}%
\emph{where }$\delta >0$\emph{\ is a sufficiently small number. Let the
number }$\varepsilon \in \left( 0,1/2\right) $\emph{\ be the one chosen in (%
\ref{3.50}). Let the function }$u\in C^{6,3}\left( \overline{G}_{T}\right)
\cap H^{4}\left( G_{T}\right) $ \emph{satisfies conditions (\ref{3.151})-(%
\ref{3.154}), where the function }$b\left( x\right) \in C^{1}\left( 
\overline{G}\right) $\emph{. In (\ref{3.151}), let the function }$R\in
C^{6,3}\left( \overline{G}_{T}\right) $\emph{\ satisfies (\ref{3.155}). Then
there exists a sufficiently small number }$\delta _{0}=\delta _{0}\left(
L,G,T,\sigma ,\varepsilon ,\nu ,A,\left\Vert R\right\Vert _{C^{6,3}\left( 
\overline{G}_{T}\right) }\right) \in \left( 0,1\right) $\emph{\ depending
only on listed parameters such that the following H\"{o}lder stability
estimates are valid:}%
\begin{equation}
\left\Vert b\right\Vert _{L_{2}\left( G_{\varepsilon }\right) }\leq
C_{1}\left( 1+\left\Vert u_{t}\right\Vert _{H^{3}\left( G_{T}\right)
}\right) \delta ^{\rho },\text{ }\forall \delta \in \left( 0,\delta
_{0}\right) ,  \label{3.20}
\end{equation}%
\begin{equation}
\left\Vert u_{t}\right\Vert _{H^{2,1}\left( G_{\varepsilon ,T}\right)
},\left\Vert u\right\Vert _{H^{2,1}\left( G_{\varepsilon ,T}\right) }\leq
C_{1}\left( 1+\left\Vert u_{t}\right\Vert _{H^{3}\left( G_{T}\right)
}\right) \delta ^{\rho },\text{ }\forall \delta \in \left( 0,\delta
_{0}\right) ,  \label{3.21}
\end{equation}%
\emph{where the numbers }$\rho $ \emph{and} $C_{1}$ \emph{depend only on
listed parameters, } \emph{\ }%
\begin{equation*}
\rho =\rho \left( L,G,T,\sigma ,\varepsilon ,\nu ,A,\left\Vert R\right\Vert
_{C^{6,3}\left( \overline{G}_{T}\right) }\right) \in \left( 0,1/2\right) ,
\end{equation*}%
\begin{equation}
C_{1}=C_{1}\left( L,G,T,\sigma ,\varepsilon ,\nu ,A,\left\Vert R\right\Vert
_{C^{6,3}\left( \overline{G}_{T}\right) }\right) >0.  \label{3.2000}
\end{equation}%
\emph{\ }

\textbf{Theorem 3} (uniqueness). \emph{Assume that conditions (\ref{1.07})-(%
\ref{1.6}), (\ref{3.1})-(\ref{3.200}) hold. Suppose that }$\delta =0$ \emph{%
in (\ref{3.18}) and (\ref{3.19})}. \emph{Then }$u\left( x,t\right) \equiv 0$%
\emph{\ in }$Q_{T}$\emph{\ and }$b\left( x\right) \equiv 0$\emph{\ in }$%
\Omega .$

We now want to avoid unnecessary technical details linked with the
evaluations of boundary terms generated by $\func{div}U$ in (\ref{3.160})
when integrating the pointwise Carleman estimate (\ref{3.15}) over the domain%
\emph{\ }$Q_{T}$ and applying Gauss formula. For this reason we restrict our
attention in Theorem 4 to the case when $\Omega $ is a rectangular prism.
Although Theorem 4 might likely be extended to the case of a more
complicated domain $\Omega ,$ this is not our goal here, see Remark 2.1.
More precisely, we assume in Theorem 4 that 
\begin{equation}
\Omega =\left\{ x:x_{1}\in \left( 0,\frac{1}{4}\right) ,\left\vert
x_{i}\right\vert <\frac{1}{2\sqrt{n-1}},i=2,...,n\right\} .  \label{3.22}
\end{equation}%
If $\Omega $ is an arbitrary rectangular prism, then the obvious linear
change of variables can transform it in (\ref{3.22}). Denote 
\begin{equation}
\partial _{1}^{+}\Omega =\left\{ x:x_{1}=\frac{1}{4},\left\vert
x_{i}\right\vert <\frac{1}{2\sqrt{n-1}},i=2,...,n\right\} \subset \partial
\Omega ,  \label{3.023}
\end{equation}%
\begin{equation}
\partial _{1}^{-}\Omega =\left\{ x:x_{1}=0,\left\vert x_{i}\right\vert <%
\frac{1}{2\sqrt{n-1}},i=2,...,n\right\} \subset \partial \Omega ,
\label{3.23}
\end{equation}%
\begin{equation}
\partial _{1}^{+}\Omega _{T}=\partial _{1}^{+}\Omega \times \left(
0,T\right) ,\text{ }\partial _{1}^{-}\Omega _{T}=\partial _{1}^{-}\Omega
\times \left( 0,T\right) .  \label{3.230}
\end{equation}%
If $n=1,$ then $\left\vert x_{i}\right\vert $ should not be parts of (\ref%
{3.22}), (\ref{3.023}) and (\ref{3.23}). In particular, $\partial
_{1}^{-}\Omega \subset \Gamma ,$ where $\Gamma $ is defined in (\ref{3.1}).
Let%
\begin{equation}
\partial _{i}^{+}\Omega =\left\{ x:x_{i}=\frac{1}{2\sqrt{n-1}}\right\} \cap
\partial \Omega ,\text{ }\partial _{i}^{+}\Omega _{T}=\partial
_{i}^{+}\Omega \times \left( 0,T\right) ,i=2,..,n,  \label{6.1}
\end{equation}%
\begin{equation}
\partial _{i}^{-}\Omega =\left\{ x:x_{i}=-\frac{1}{2\sqrt{n-1}}\right\} \cap
\partial \Omega ,\text{ }\partial _{i}^{-}\Omega _{T}=\partial
_{i}^{-}\Omega \times \left( 0,T\right) ,i=2,..,n.  \label{6.2}
\end{equation}%
Using (\ref{1.07}), (\ref{3.22})-(\ref{6.2}), we obtain%
\begin{equation}
\partial \Omega =\left( \cup _{i=1}^{n}\partial _{i}^{+}\Omega \right) \cup
\left( \cup _{i=1}^{n}\partial _{i}^{-}\Omega \right) ,  \label{6.3}
\end{equation}%
\begin{equation}
S_{T}=\left( \cup _{i=1}^{n}\partial _{i}^{+}\Omega _{T}\right) \cup \left(
\cup _{i=1}^{n}\partial _{i}^{-}\Omega _{T}\right) .  \label{6.4}
\end{equation}%
It follows from (\ref{6.4}) that $S_{T}$ is not smooth. On the other hand,
we need the norm of the space $H^{k,0}\left( S_{T}\right) $ in Theorem 4.
Hence, using (\ref{3.23})-(\ref{6.4}), we define this space as%
\begin{equation}
H^{k,0}\left( S_{T}\right) =\left\{ 
\begin{array}{c}
s\left( x,t\right) : \\ 
s\in H^{k,0}\left( \partial _{i}^{+}\Omega _{T}\right) ,s\in H^{k,0}\left(
\partial _{i}^{-}\Omega _{T}\right) ,i=1,...,n, \\ 
\left\Vert s\right\Vert _{H^{k,0}\left( S_{T}\right) }^{2}= \\ 
+\dsum\limits_{i=1}^{n}\left( \left\Vert s\right\Vert _{H^{k,0}\left(
\partial _{i}^{+}\Omega _{T}\right) }^{2}+\left\Vert s\right\Vert
_{H^{k,0}\left( \partial _{i}^{-}\Omega _{T}\right) }^{2}\right)%
\end{array}%
\right\} ,k=1,2.  \label{6.40}
\end{equation}

\textbf{Theorem 4} (Lipschitz stability). \emph{Assume that conditions (\ref%
{1.07})-(\ref{1.5}), (\ref{3.1})-(\ref{3.4}) hold. Let the function }$u\in
C^{6,3}\left( \overline{Q}_{T}\right) $\emph{\ satisfies conditions (\ref%
{3.151})-(\ref{3.153}), in which the domain }$G$\emph{\ is replaced with the
domain }$\Omega $\emph{\ defined in (\ref{3.22}). Assume that the Dirichlet
and Neumann boundary conditions are given on the entire \ lateral boundary }$%
S_{T},$ \emph{i.e. we assume that (\ref{3.154}) is replaced with}%
\begin{equation}
u\mid _{S_{T}}=p\left( x,t\right) ,\text{ }\partial _{n}u\mid
_{S_{T}}=q\left( x,t\right) .  \label{3.24}
\end{equation}%
\emph{Let the function }$b\left( x\right) \in C^{1}\left( \overline{\Omega }%
\right) $\emph{.\ Let in (\ref{3.151}) the function }$R\in C^{6,3}\left( 
\overline{Q}_{T}\right) $\emph{\ and let inequality (\ref{1.16}) be valid}$.$%
\emph{\ Then the following Lipschitz stability estimates hold:} 
\begin{equation}
\left\Vert b\right\Vert _{L_{2}\left( \Omega \right) }\leq C_{2}\left(
\left\Vert p_{t}\right\Vert _{H^{2,0}\left( S_{T}\right) }+\left\Vert
q_{t}\right\Vert _{H^{1,0}\left( S_{T}\right) }+\left\Vert f\right\Vert
_{H^{4}\left( \Omega \right) }+\left\Vert F\right\Vert _{H^{4}\left( \Omega
\right) }\right) ,  \label{3.25}
\end{equation}%
\begin{equation*}
\left\Vert u\right\Vert _{H^{2,1}\left( Q_{T}\right) },\left\Vert
u_{t}\right\Vert _{H^{2,1}\left( Q_{T}\right) }\leq
\end{equation*}%
\begin{equation}
\leq C_{2}\left( \left\Vert p_{t}\right\Vert _{H^{2,0}\left( S_{T}\right)
}+\left\Vert q_{t}\right\Vert _{_{H^{1,0}\left( S_{T}\right) }}+\left\Vert
f\right\Vert _{H^{4}\left( \Omega \right) }+\left\Vert F\right\Vert
_{H^{4}\left( \Omega \right) }\right) ,  \label{3.26}
\end{equation}%
\emph{where the number }%
\begin{equation}
C_{2}=C_{2}\left( L,\Omega ,T,\sigma ,\nu ,A,\left\Vert R\right\Vert
_{C^{6,3}\left( \overline{G}_{T}\right) }\right) >0  \label{3.27}
\end{equation}%
\emph{depends only on listed parameters.}

\section{Proof of Theorem 2}

\label{sec:5}

In this section $\left( x,t\right) \in G_{T}$ and $C_{1}>0$\ denotes
different positive numbers depending only on parameters listed in (\ref%
{3.2000}).\emph{\ }The function $w\left( x,t\right) ,$ which we introduce
below in this section, is not the one we have used in the proof of Theorem 1.

Divide both sides of equation (\ref{3.151}) by $R\left( x,t\right) ,$ which
we can do by (\ref{3.155}). Denote%
\begin{equation}
v\left( x,t\right) =\frac{u\left( x,t\right) }{R\left( x,t\right) },\text{ }%
\widetilde{f}\left( x\right) =\frac{f\left( x\right) }{R\left( x,0\right) },%
\text{ }\widetilde{F}\left( x\right) =\frac{F\left( x\right) }{R\left(
x,T\right) },\text{ }\left( x,t\right) \in G_{T},  \label{5.1}
\end{equation}%
\begin{equation}
\widetilde{p}\left( x,t\right) =\frac{p\left( x,t\right) }{R\left(
x,t\right) },\text{ }\left( x,t\right) \in \Gamma _{T},  \label{5.2}
\end{equation}%
\begin{equation}
\text{ }\widetilde{q}\left( x,t\right) =\frac{q\left( x,t\right) }{R\left(
x,t\right) }-p\left( x,t\right) \frac{\partial _{n}R\left( x,t\right) }{%
R^{2}\left( x,t\right) },\left( x,t\right) \in \Gamma _{T}.  \label{5.02}
\end{equation}%
It follows from (\ref{3.155}) and (\ref{5.02}) that 
\begin{equation}
\left\Vert \widetilde{p}_{t}\right\Vert _{H^{2,0}\left( \Gamma _{T}\right)
}\leq C_{1}\left\Vert p_{t}\right\Vert _{H^{2,0}\left( \Gamma _{T}\right) },%
\text{ }\left\Vert \widetilde{q}_{t}\right\Vert _{H^{2,0}\left( \Gamma
_{T}\right) }\leq C_{1}\left\Vert q_{t}\right\Vert _{H^{2,0}\left( \Gamma
_{T}\right) }.  \label{5.002}
\end{equation}%
Then (\ref{3.151})-(\ref{3.154}) become:%
\begin{equation}
v_{t}=\widetilde{L}v+b\left( x\right) \text{ in }G_{T},  \label{5.3}
\end{equation}%
\begin{equation}
v\left( x,0\right) =\widetilde{f}\left( x\right) \text{ in }G,  \label{5.4}
\end{equation}%
\begin{equation}
v\left( x,T\right) =\widetilde{F}\left( x\right) \text{ in }G,  \label{5.5}
\end{equation}%
\begin{equation}
v\mid _{\Gamma _{T}}=\widetilde{p}\left( x,t\right) ,\partial _{n}v\mid
_{\Gamma _{T}}=\widetilde{q}\left( x,t\right) ,  \label{5.6}
\end{equation}%
where $\widetilde{L}$ is the operator, which is obtained from the operator $%
L $ in the obvious way, and by (\ref{1.5}) and (\ref{1.6}) 
\begin{equation}
\widetilde{L}v=L_{0}v+\widetilde{L}_{1}v,  \label{5.7}
\end{equation}%
where the principal part $L_{0}v$ of $\widetilde{L}v$ is defined in (\ref%
{1.5}) and $\widetilde{L}_{1}v$ contains only lower order derivatives of the
function $v$. By (\ref{5.3})-(\ref{5.6})%
\begin{equation}
v_{t}\left( x,0\right) =\widetilde{L}\left( \widetilde{f}\left( x\right)
\right) +b\left( x\right) ,v_{t}\left( x,T\right) =\widetilde{L}\left( 
\widetilde{F}\left( x\right) \right) +b\left( x\right) .  \label{5.8}
\end{equation}%
Introduce a new function $w\left( x,t\right) ,$%
\begin{equation}
w\left( x,t\right) =\partial _{t}v\left( x,t\right) -\left( \frac{t}{T}%
\widetilde{L}\left( \widetilde{F}\left( x\right) \right) +\left( 1-\frac{t}{T%
}\right) \widetilde{L}\left( \widetilde{f}\left( x\right) \right) \right) .
\label{5.9}
\end{equation}%
Then $w\in C^{4,2}\left( \overline{G}_{T}\right) \cap H^{3}\left(
G_{T}\right) .$ Also, (\ref{5.3})-(\ref{5.9}) imply%
\begin{equation}
w_{t}=\widetilde{L}w+\left( \partial _{t}\widetilde{L}_{1}\right) v+P\left(
x,t\right) ,  \label{5.10}
\end{equation}%
\begin{equation}
w\mid _{\Gamma _{T}}=\overline{p}_{t}\left( x,t\right) ,\partial _{n}w\mid
_{_{\Gamma _{T}}}=\overline{q}_{t}\left( x,t\right) ,  \label{5.11}
\end{equation}%
\begin{equation}
w\left( x,0\right) =b\left( x\right) ,w\left( x,T\right) =b\left( x\right) ,
\label{5.12}
\end{equation}%
where $\partial _{t}\widetilde{L}_{1}$ means that $t-$dependent coefficients
of the operator $\widetilde{L}_{1}$ are differentiated once with respect to $%
t.$ In (\ref{5.10}) and (\ref{5.11})%
\begin{equation}
\left. 
\begin{array}{c}
P\left( x,t\right) = \\ 
=\left[ \widetilde{L}\left( \widetilde{f}\left( x\right) -\widetilde{F}%
\left( x\right) \right) +t\widetilde{L}^{2}\left( \widetilde{F}\left(
x\right) \right) +\left( T-t\right) \widetilde{L}^{2}\left( \widetilde{f}%
\left( x\right) \right) \right] /T,\text{ }\left( x,t\right) \in G_{T},%
\end{array}%
\right.  \label{5.13}
\end{equation}
\begin{equation*}
P\left( x,t\right) =
\end{equation*}%
\begin{equation}
\overline{p}\left( x,t\right) =\partial _{t}\widetilde{p}\left( x,t\right)
-\left( \frac{t}{T}\widetilde{L}\left( \widetilde{F}\left( x\right) \right)
+\left( 1-\frac{t}{T}\right) \widetilde{L}\left( \widetilde{f}\left(
x\right) \right) \right) ,\text{ }\left( x,t\right) \in \Gamma _{T},
\label{5.14}
\end{equation}%
\begin{equation}
\overline{q}\left( x,t\right) =\partial _{t}\widetilde{q}\left( x,t\right)
-\partial _{n}\left( \frac{t}{T}\widetilde{L}\left( \widetilde{F}\left(
x\right) \right) +\left( 1-\frac{t}{T}\right) \widetilde{L}\left( \widetilde{%
f}\left( x\right) \right) \right) ,\text{ }\left( x,t\right) \in \Gamma _{T}.
\label{5.15}
\end{equation}%
Also, by (\ref{5.4}) and (\ref{5.9})%
\begin{equation}
\left. 
\begin{array}{c}
v\left( x,t\right) =\dint\limits_{0}^{t}w\left( x,\tau \right) d\tau +%
\widetilde{f}\left( x\right) + \\ 
+\dint\limits_{0}^{t}\left[ \left( \tau /T\right) \widetilde{L}\left( 
\widetilde{F}\left( x\right) \right) +\left( 1-\left( \tau /T\right) \right) 
\widetilde{L}\left( \widetilde{f}\left( x\right) \right) \right] d\tau .%
\end{array}%
\right.  \label{5.16}
\end{equation}%
Substituting (\ref{5.9}) in (\ref{5.10}), making the resulting equation
stronger by replacing it with the inequality and using (\ref{3.155}), (\ref%
{5.1})-(\ref{5.02}) and (\ref{5.13}), we obtain%
\begin{equation}
\left\vert w_{t}-L_{0}w\right\vert \leq C_{1}\left( \left\vert \nabla
w\right\vert +\left\vert w\right\vert +\dint\limits_{0}^{t}\left( \left\vert
\nabla w\right\vert +\left\vert w\right\vert \right) \left( x,\tau \right)
d\tau \right) +K\left( x,t\right) ,\left( x,t\right) \in G_{T},  \label{5.17}
\end{equation}%
where the function $K\left( x,t\right) \geq 0,K\in L_{2}\left( G_{T}\right) $
and is such that%
\begin{equation}
\left\Vert K\right\Vert _{L_{2}\left( G_{T}\right) }\leq C_{1}\left(
\left\Vert f\right\Vert _{H^{4}\left( G\right) }+\left\Vert F\right\Vert
_{H^{4}\left( G\right) }\right) .  \label{5.18}
\end{equation}

We are ready now to apply Theorem 1 to inequality (\ref{5.17}), which is
supplied by conditions (\ref{5.11}) and (\ref{5.12}). Since the function $%
\phi =\phi \left( x\right) $ is independent on $t$, then 
\begin{equation}
\dint\limits_{G_{T}}\left( \dint\limits_{0}^{t}\left( \left\vert \nabla
w\right\vert +\left\vert w\right\vert \right) \left( x,\tau \right) d\tau
\right) ^{2}\phi ^{2}\leq C_{1}\dint\limits_{G_{T}}\left( \left\vert \nabla
w\right\vert ^{2}+w^{2}\right) \phi ^{2}dxdt.  \label{5.19}
\end{equation}%
Square both sides of (\ref{5.17}), multiply by the function $\phi ^{2}$ with 
$\mu =\mu _{0}$ and integrate over the domain $G_{T}.$ Using (\ref{5.19})
and Cauchy-Schwarz inequality, we obtain%
\begin{equation}
\dint\limits_{G_{T}}\left( w_{t}-L_{0}w\right) ^{2}\phi ^{2}dxdt\leq
C_{1}\dint\limits_{G_{T}}\left( \left\vert \nabla w\right\vert
^{2}+w^{2}\right) \phi ^{2}dxdt+C_{1}\dint\limits_{G_{T}}K^{2}\phi ^{2}dxdt.
\label{5.20}
\end{equation}%
Integrate the pointwise Carleman estimate (\ref{3.15}) of Theorem 1 with $%
\mu =\mu _{0}$ over the domain $G_{T}$ and use (\ref{3.5})-(\ref{3.8}), (\ref%
{3.16}), (\ref{3.160}) and Gauss formula. Next, apply the resulting estimate
to the left hand side of (\ref{5.20}) for all $\lambda \geq \lambda _{0}$.
Using (\ref{3.1}), (\ref{3.200}), (\ref{3.7}), (\ref{3.8}) and (\ref{5.20}),
we obtain%
\begin{equation*}
C_{1}\dint\limits_{G_{T}}\left( \left\vert \nabla w\right\vert
^{2}+w^{2}\right) \phi ^{2}dxdt+C_{1}\dint\limits_{G_{T}}K^{2}\phi
^{2}dxdt\geq
\end{equation*}%
\begin{equation}
\geq \frac{C}{\lambda }\dint\limits_{G_{T}}\left(
w_{t}^{2}+\dsum\limits_{i,j=1}^{n}w_{x_{i}x_{j}}^{2}\right) \phi
^{2}dxdt+C\dint\limits_{G_{T}}\left( \lambda \left( \nabla w\right)
^{2}+\lambda ^{3}w^{2}\right) \phi ^{2}dxdt+  \label{5.21}
\end{equation}%
\begin{equation*}
+\dint\limits_{\Gamma _{T}}U\left( 0,\overline{x},t\right) \cdot n_{\Gamma }d%
\overline{x}dt+\dint\limits_{\partial _{2}G_{T}}U\cdot ndSdt+
\end{equation*}%
\begin{equation*}
+\dint\limits_{G_{T}}\partial _{t}Vdxdt,\text{ }\forall \lambda \geq \lambda
_{0},
\end{equation*}%
where $a\cdot b$ denotes the scalar product in $\mathbb{R}^{n}$ of any two
vectors $a,b\in $ $\mathbb{R}^{n}.$ Further, vectors%
\begin{equation}
n_{\Gamma }=\left( -1,0,...,0\right) ^{T}  \label{5.210}
\end{equation}%
and $n$ are outward looking unit normal vectors at $\Gamma _{T}$ and $%
\partial _{2}G$ respectively.

It follows from (\ref{3.16}) and (\ref{5.12}) that the condition of
Corollary is valid, i.e.%
\begin{equation*}
\left. 
\begin{array}{c}
V\left( x,T\right) =V\left( x,0\right) = \\ 
=\left( d/2\right) \dsum\limits_{i,j=1}^{n}\left[ a^{ij}\varphi ^{\mu
+2}\left( b_{i}-\lambda \mu _{0}\varphi _{i}\varphi ^{-\mu _{0}-1}b\right)
\left( b_{j}-\lambda \mu _{0}\varphi _{j}\varphi ^{-\mu _{0}-1}b\right) \phi
^{2}\right] \left( x\right) + \\ 
-\left( d/2\right) \left[ \lambda ^{2}\mu _{0}^{2}\varphi ^{-\mu
_{0}}\dsum\limits_{i,j=1}^{n}a^{ij}\left( x\right) \left( \varphi
_{i}\varphi _{j}\left( 1-\lambda ^{-1}\left( 1+\mu _{0}^{-1}\right) \varphi
^{\mu }\right) \right) b^{2}\phi ^{2}\right] \left( x\right) + \\ 
+\left( d/2\right) \left[ \lambda \mu _{0}b^{2}\phi ^{2}+\left( 4^{2\mu
_{0}+2}\left( \lambda \mu _{0}\right) \right)
^{-1}\dsum\limits_{i,j=1}^{n}a^{ij}\left( x\right) b_{i}b_{j}\phi ^{2}\right]
\left( x\right) .%
\end{array}%
\right.
\end{equation*}

Hence, 
\begin{equation}
\dint\limits_{G_{T}}\partial _{t}Vdxdt=0.  \label{5.23}
\end{equation}

We now analyze which norms of functions $p_{t}$ and $q_{t}$ should be
included in the estimate of the integral 
\begin{equation}
\dint\limits_{\Gamma _{T}}U\left( 0,\overline{x},t\right) \cdot n_{\Gamma }d%
\overline{x}dt  \label{1}
\end{equation}%
in (\ref{5.21}). Consider the term%
\begin{equation}
B_{ijks}=d\left( 4^{2\mu _{0}+2}\left( \lambda \mu _{0}\right) \right) ^{-1} 
\left[ \left( a^{ij}\left( x\right) a^{ks}\left( x\right) w_{i}w_{ks}\phi
^{2}\right) _{j}+\left( -a^{ij}\left( x\right) a^{ks}\left( x\right)
w_{i}w_{sj}\phi ^{2}\right) _{k}\right]  \label{2}
\end{equation}%
in the last two lines of (\ref{3.160}), where $u$ is replaced with $w$. It
follows from (\ref{3.1})-(\ref{3.200}), (\ref{3.7}), (\ref{5.210}) and (\ref%
{2}) that only the following cases can provide a non-zero impact in integral
(\ref{1}):

\begin{equation}
j=k=1,  \label{3}
\end{equation}%
\begin{equation}
j=1,k\neq 1,  \label{30}
\end{equation}%
\begin{equation}
j\neq 1,\text{ }k=1.  \label{31}
\end{equation}%
In the case (\ref{3}) we obtain from (\ref{2}):%
\begin{equation*}
\left. 
\begin{array}{c}
B_{ijks}=B_{i11s}= \\ 
=d\left( 4^{2\mu _{0}+2}\left( \lambda \mu _{0}\right) \right) ^{-1}\left[
\left( a^{i1}\left( x\right) a^{1s}\left( x\right) w_{i}w_{1s}\phi
^{2}\right) _{1}+\left( -a^{i1}\left( x\right) a^{1s}\left( x\right)
w_{i}w_{1s}\phi ^{2}\right) _{1}\right] =0.%
\end{array}%
\right.
\end{equation*}%
In the case (\ref{30}) the second term of $B_{ijks}$ provides zero impact in
integral (\ref{1}). Similarly, in the case (\ref{31}) the first term of $%
B_{ijks}$ provides zero impact in integral (\ref{1}). Hence, we should
include norms $\left\Vert \overline{p}_{t}\right\Vert _{H^{2,0}\left( \Gamma
_{T}\right) }$ and $\left\Vert \overline{q}_{t}\right\Vert _{H^{1,0}\left(
\Gamma _{T}\right) }$ in the estimate of integral (\ref{1}). Hence, using (%
\ref{3.160}), (\ref{3.155}), (\ref{5.1})-(\ref{5.02}), (\ref{5.11}), (\ref%
{5.14}) and (\ref{5.15}), we obtain the following estimate from the below of
the integral in (\ref{1}): 
\begin{equation*}
\dint\limits_{\Gamma _{T}}U\left( 0,\overline{x},t\right) \cdot n_{\Gamma }d%
\overline{x}dt\geq -C_{1}\lambda ^{3}\left( \max_{\overline{G}}\phi
^{2}\right) \left( \left\Vert \overline{p}_{t}\right\Vert _{H^{2,0}\left(
\Gamma _{T}\right) }^{2}+\left\Vert \overline{q}_{t}\right\Vert
_{H^{1,0}\left( \Gamma _{T}\right) }^{2}\right) =
\end{equation*}%
\begin{equation}
=-C_{1}\lambda ^{3}\exp \left( 2\lambda \left( \frac{1}{4}\right) ^{-\mu
_{0}}\right) \left( \left\Vert \overline{p}_{t}\right\Vert _{H^{2,0}\left(
\Gamma _{T}\right) }^{2}+\left\Vert \overline{q}_{t}\right\Vert
_{H^{1,0}\left( \Gamma _{T}\right) }^{2}\right) ,\text{ }\forall \lambda
\geq \lambda _{0}.  \label{5.24}
\end{equation}%
Next, it follows from (\ref{3.4}), (\ref{3.8}), (\ref{3.13}), (\ref{3.160})
and the trace theorem that the second term in the third line of (\ref{5.21})
can be estimated as: 
\begin{equation*}
\dint\limits_{\partial _{2}G_{T}}\left( U\cdot n\right) dSdt\geq
\end{equation*}%
\begin{equation}
\geq -C_{1}\lambda ^{3}\exp \left( 2\lambda \left( \frac{3}{4}\right) ^{-\mu
_{0}}\right) \dint\limits_{\partial _{2}G_{T}}\left(
\dsum\limits_{i,j=1}^{n}w_{ij}^{2}+w_{t}^{2}+\left( \nabla w\right)
^{2}+w^{2}\right) dSdt\geq  \label{5.25}
\end{equation}%
\begin{equation*}
\geq -C_{1}\lambda ^{3}\exp \left( 2\lambda \left( \frac{3}{4}\right) ^{-\mu
_{0}}\right) \left\Vert w\right\Vert _{H^{3}\left( G_{T}\right) }^{2},\text{ 
}\forall \lambda \geq \lambda _{0}.
\end{equation*}%
It follows from (\ref{5.002}), (\ref{5.14}) and (\ref{5.15}) that%
\begin{equation}
\left. 
\begin{array}{c}
\left\Vert \overline{p}_{t}\right\Vert _{H^{2,0}\left( \Gamma _{T}\right)
}^{2}+\left\Vert \overline{q}_{t}\right\Vert _{H^{1,0}\left( \Gamma
_{T}\right) }^{2}\leq \\ 
\leq C_{1}\left( \left\Vert p_{t}\right\Vert _{H^{2,0}\left( \Gamma
_{T}\right) }^{2}+\left\Vert q_{t}\right\Vert _{H^{1,0}\left( \Gamma
_{T}\right) }^{2}+\left\Vert f\right\Vert _{H^{4}\left( G\right)
}^{2}+\left\Vert F\right\Vert _{H^{4}\left( G\right) }^{2}\right) .%
\end{array}%
\right.  \label{5.250}
\end{equation}%
Combining (\ref{5.21}) with (\ref{5.18}), (\ref{5.23}), (\ref{5.24}), (\ref%
{5.25} and (\ref{5.250}), we obtain%
\begin{equation*}
C_{1}\lambda ^{3}\exp \left( 2\lambda \left( \frac{1}{4}\right) ^{-\mu
_{0}}\right) \left( \left\Vert p_{t}\right\Vert _{H^{2,0}\left( \Gamma
_{T}\right) }^{2}+\left\Vert q_{t}\right\Vert _{H^{1,0}\left( \Gamma
_{T}\right) }^{2}+\left\Vert f\right\Vert _{H^{4}\left( G\right)
}^{2}+\left\Vert F\right\Vert _{H^{4}\left( G\right) }^{2}\right) +
\end{equation*}%
\begin{equation*}
+C_{1}\lambda ^{3}\exp \left( 2\lambda \left( \frac{3}{4}\right) ^{-\mu
_{0}}\right) \left\Vert w\right\Vert _{H^{3}\left( G_{T}\right)
}^{2}+C_{1}\dint\limits_{G_{T}}\left( \left\vert \nabla w\right\vert
^{2}+w^{2}\right) \phi ^{2}dxdt\geq
\end{equation*}%
\begin{equation}
\geq \frac{1}{\lambda }\dint\limits_{G_{T}}\left(
w_{t}^{2}+\dsum\limits_{i,j=1}^{n}w_{x_{i}x_{j}}^{2}\right) \phi
^{2}dxdt+\dint\limits_{G_{T}}\left( \lambda \left( \nabla w\right)
^{2}+\lambda ^{3}w^{2}\right) \phi ^{2}dxdt.  \label{5.26}
\end{equation}%
Choose $\lambda _{1}=\lambda _{1}\left( L,G,T,\sigma ,\nu ,A,\left\Vert
R\right\Vert _{C^{6,3}\left( \overline{G}_{T}\right) }\right) \geq \lambda
_{0}\geq 1$ so large that $C_{1}<\lambda _{1}/2.$ Then (\ref{5.26}) becomes%
\begin{equation*}
C_{1}\lambda ^{2}\exp \left( 2\lambda \left( \frac{1}{4}\right) ^{-\mu
_{0}}\right) \left( \left\Vert p_{t}\right\Vert _{H^{2,0}\left( \Gamma
_{T}\right) }^{2}+\left\Vert q_{t}\right\Vert _{H^{1,0}\left( \Gamma
_{T}\right) }^{2}+\left\Vert f\right\Vert _{H^{4}\left( G\right)
}^{2}+\left\Vert F\right\Vert _{H^{4}\left( G\right) }^{2}\right) +
\end{equation*}%
\begin{equation}
+C_{1}\lambda ^{3}\exp \left( 2\lambda \left( \frac{3}{4}\right) ^{-\mu
_{0}}\right) \left\Vert w\right\Vert _{H^{3}\left( G_{T}\right) }^{2}\geq
\label{5.27}
\end{equation}%
\begin{equation*}
\geq \frac{1}{\lambda }\dint\limits_{G_{T}}\left(
w_{t}^{2}+\dsum\limits_{i,j=1}^{n}w_{x_{i}x_{j}}^{2}\right) \phi
^{2}dxdt+\dint\limits_{G_{T}}\left( \lambda \left( \nabla w\right)
^{2}+\lambda ^{3}w^{2}\right) \phi ^{2}dxdt,\forall \lambda \geq \lambda
_{1}.
\end{equation*}%
Replace in the last line of (\ref{5.27}) $G_{T}$ with $G_{\varepsilon
,T}\subset G_{T},$ where the domain $G_{\varepsilon ,T}$ was defined in (\ref%
{3.06}) and (\ref{3.60}). Using (\ref{3.12}), we obtain 
\begin{equation*}
\left\Vert w\right\Vert _{H^{2,1}\left( G_{\varepsilon ,T}\right) }^{2}\leq
\end{equation*}%
\begin{equation*}
\leq C_{1}\exp \left( 3\lambda \left( \frac{1}{4}\right) ^{-\mu _{0}}\right)
\left( \left\Vert p_{t}\right\Vert _{H^{2,0}\left( \Gamma _{T}\right)
}^{2}+\left\Vert q_{t}\right\Vert _{H^{1,0}\left( \Gamma _{T}\right)
}^{2}+\left\Vert f\right\Vert _{H^{4}\left( G\right) }^{2}+\left\Vert
F\right\Vert _{H^{4}\left( G\right) }^{2}\right) +
\end{equation*}%
\begin{equation}
+C_{1}\exp \left[ -\lambda \left( \left( \frac{3}{4}-\varepsilon \right)
^{-\mu _{0}}-\left( \frac{3}{4}\right) ^{-\mu _{0}}\right) \right]
\left\Vert w\right\Vert _{H^{3}\left( G_{T}\right) }^{2},\text{ }\forall
\lambda \geq \lambda _{1}.  \label{5.28}
\end{equation}

Consider the second line of (\ref{5.28}). By (\ref{3.18}), (\ref{3.19}), (%
\ref{5.1})- (\ref{5.02}), (\ref{5.11}), (\ref{5.14}) and (\ref{5.15})%
\begin{equation}
\left\Vert p_{t}\right\Vert _{H^{2,0}\left( \Gamma _{T}\right)
}^{2}+\left\Vert q_{t}\right\Vert _{H^{1,0}\left( \Gamma _{T}\right)
}^{2}+\left\Vert f\right\Vert _{H^{4}\left( G\right) }^{2}+\left\Vert
F\right\Vert _{H^{4}\left( G\right) }^{2}\leq C_{1}\delta ^{2}.  \label{5.29}
\end{equation}%
Choose the number $\delta _{0}=\delta _{0}\left( L,G,T,\sigma ,\nu
,A,\left\Vert R\right\Vert _{C^{6,3}\left( \overline{G}_{T}\right) }\right)
\in \left( 0,1\right) $ so small that%
\begin{equation*}
\exp \left( 3\lambda _{1}\left( \frac{1}{4}\right) ^{-\mu _{0}}\right) =%
\frac{1}{\delta _{0}}.
\end{equation*}%
Hence, 
\begin{equation*}
\lambda _{1}=\ln \left( \delta _{0}^{-1/\left( 3\cdot 4^{\mu _{0}}\right)
}\right) .
\end{equation*}%
Hence, 
\begin{equation}
\left. 
\begin{array}{c}
\exp \left( 3\cdot 4^{\mu _{0}}\lambda \right) \delta ^{2}=\delta ,\text{ }
\\ 
\lambda =\lambda \left( \delta \right) =\ln \left( \delta ^{-1/\left( 3\cdot
4^{\mu _{0}}\right) }\right) >\lambda _{1},\text{ }\forall \delta \in \left(
0,\delta _{0}\right) .%
\end{array}%
\right.  \label{5.30}
\end{equation}%
Hence, 
\begin{equation}
\exp \left[ -\lambda \left( \delta \right) \left( \left( \frac{3}{4}%
-\varepsilon \right) ^{-\mu _{0}}-\left( \frac{3}{4}\right) ^{-\mu
_{0}}\right) \right] =\delta ^{2\rho }.  \label{5.31}
\end{equation}%
It follows from (\ref{3.50}) and (\ref{5.30}) that in (\ref{5.31}) the
number $\rho $ is such that 
\begin{equation*}
\rho =\rho \left( L,G,T,\sigma ,\varepsilon ,\nu ,A,\left\Vert R\right\Vert
_{C^{6,3}\left( \overline{G}_{T}\right) }\right) \in \left( 0,1/2\right) .
\end{equation*}%
Hence, setting in (\ref{5.28}) $\lambda =\lambda \left( \delta \right) $, we
obtain%
\begin{equation}
\left\Vert w\right\Vert _{H^{2,1}\left( G_{\varepsilon ,T}\right) }\leq
C_{1}\left( 1+\left\Vert w\right\Vert _{H^{3}\left( G_{T}\right) }\right)
\delta ^{\rho },\text{ }\forall \delta \in \left( 0,\delta _{0}\right) .
\label{5.32}
\end{equation}%
Returning in (\ref{5.32}) from $w$ to $u$ via (\ref{5.1}) and (\ref{5.16})
and using (\ref{3.155}), and (\ref{5.12}), we obtain the target estimates (%
\ref{3.20}) and (\ref{3.21}). $\square $

\section{Proof of Theorem 3}

\label{sec:6}

Since $\delta =0$ in (\ref{3.18}) and (\ref{3.19}), then (\ref{3.20}) and (%
\ref{3.21}) imply that $u\left( x,t\right) =0$ in $G_{\varepsilon ,T}$ and $%
b\left( x\right) =0$ in $G_{\varepsilon }.$ Setting $\varepsilon \rightarrow
0,$ we obtain $u\left( x,t\right) =0$ in $G_{T}$ and $b\left( x\right) =0$
in $G.$ Changing coordinates in $\mathbb{R}^{n}$ via linear transformations,
we can sequentially cover the entire domain $\Omega $ by a sequence $\left\{
G_{k}\right\} _{k=0}^{m}$ of $G-$like subdomains, where $G_{0}=G$. This
sequence can be arranged in such a way that each intersection $G_{k+1}\cap
G_{k}$ has its sub-boundary the hypersurface like the hypersurface $\Gamma $
in (\ref{3.1}), (\ref{3.200}). Thus, if $u\left( x,t\right) =0$ in $%
G_{k}\times \left( 0,T\right) $ and $b\left( x\right) =0$ in $G_{k},$ then
Theorem 2 implies that $u\left( x,t\right) =0$ in $G_{k+1}\times \left(
0,T\right) $ and $b\left( x\right) =0$ in $G_{k+1}$ as well. Thus, $u\left(
x,t\right) \equiv 0$ in $Q_{T}$ and $b\left( x\right) \equiv 0$ in $\Omega .$
$\square $

\section{Proof of Theorem 4}

\label{sec:7}

In this section $\left( x,t\right) \in Q_{T}$ and $C_{2}>0$\ denotes
different positive numbers depending only on parameters listed in (\ref{3.27}%
). Recall that the domain $\Omega $ is the one defined in (\ref{3.22}),
also, see (\ref{3.023})-(\ref{6.4}). We now keep the same notations as the
ones in the proof of Theorem 2 with the only obvious changes of $G$ and $%
G_{T}$ with $\Omega $ and $Q_{T}$ respectively as well as those changes,
which are generated by (\ref{3.22})-(\ref{6.4}).

Using (\ref{3.023})-(\ref{6.4}), we obtain similarly with (\ref{5.21})%
\begin{equation*}
C_{2}\dint\limits_{Q_{T}}\left( \left\vert \nabla w\right\vert
^{2}+w^{2}\right) \phi ^{2}dxdt+C_{2}\dint\limits_{Q_{T}}K^{2}\phi
^{2}dxdt\geq
\end{equation*}%
\begin{equation*}
\geq \frac{C}{\lambda }\dint\limits_{Q_{T}}\left(
w_{t}^{2}+\dsum\limits_{i,j=1}^{n}w_{x_{i}x_{j}}^{2}\right) \phi
^{2}dxdt+C\dint\limits_{Q_{T}}\left( \lambda \left( \nabla w\right)
^{2}+\lambda ^{3}w^{2}\right) \phi ^{2}dxdt+
\end{equation*}%
\begin{equation}
+\dsum\limits_{i=1}^{n}\dint\limits_{\partial _{i}^{+}\Omega _{T}}U\cdot
n_{i}dSdt-\dsum\limits_{i=1}^{n}\dint\limits_{\partial _{i}^{-}\Omega
_{T}}\left( U\cdot n_{i}\right) dSdt+\dint\limits_{Q_{T}}\partial _{t}Vdxdt,%
\text{ }\forall \lambda \geq \lambda _{0},  \label{6.5}
\end{equation}%
where $\lambda _{0}$ was chosen in Theorem 1. In (\ref{6.5}), $n_{i}=\left(
0,..,1,0...0\right) ^{T},$ where \textquotedblleft $1$" is the component
number $i$. The vector function $U$ is the same as in (\ref{3.160}), in
which $u$ is replaced with $w$. The key equality 
\begin{equation}
\dint\limits_{Q_{T}}\partial _{t}Vdxdt=0  \label{6.6}
\end{equation}%
is proven completely similarly with (\ref{5.23}). As to the function $%
K\left( x,t\right) $ in (\ref{6.5}), similarly with (\ref{5.18}) 
\begin{equation}
\left\Vert K\right\Vert _{L_{2}\left( Q_{T}\right) }\leq C_{2}\left(
\left\Vert f\right\Vert _{H^{4}\left( \Omega \right) }+\left\Vert
F\right\Vert _{H^{4}\left( \Omega \right) }\right) .  \label{6.7}
\end{equation}

Using (\ref{3.160}), we obtain completely similarly with (\ref{5.24})%
\begin{equation*}
\dsum\limits_{i=1}^{n}\dint\limits_{\partial _{i}^{+}\Omega _{T}}U\cdot
n_{i}dSdt-\dsum\limits_{i=1}^{n}\dint\limits_{\partial _{i}^{-}\Omega
_{T}}U\cdot n_{i}dSdt\geq
\end{equation*}%
\begin{equation}
\geq -C_{2}\lambda ^{3}\exp \left( 2\lambda \left( \frac{1}{4}\right) ^{-\mu
_{0}}\right) \left( \left\Vert \overline{p}_{t}\right\Vert _{H^{2,0}\left(
S_{T}\right) }^{2}+\left\Vert \overline{q}_{t}\right\Vert _{H^{1,0}\left(
S_{T}\right) }^{2}\right) ,\text{ }\forall \lambda \geq \lambda _{0},
\label{6.8}
\end{equation}%
where notations \ref{5.14}) and (\ref{5.15}) are kept with the replacement
of $\Gamma _{T}$ with $S_{T}.$ Choose $\lambda _{2}=\lambda _{2}\left(
L,\Omega ,T,\sigma ,\nu ,A,\left\Vert R\right\Vert _{C^{6,3}\left( \overline{%
G}_{T}\right) }\right) \geq \lambda _{1}\geq 1$ so large that $C_{2}<\lambda
_{2}/2.$ Using (\ref{6.5})-(\ref{6.8}) and the obvious analog of (\ref{5.250}%
), we obtain%
\begin{equation*}
C_{2}\exp \left( 3\lambda \left( \frac{1}{4}\right) ^{-\mu _{0}}\right)
\left( \left\Vert p_{t}\right\Vert _{H^{2,0}\left( S_{T}\right)
}^{2}+\left\Vert q_{t}\right\Vert _{H^{1,0}\left( S_{T}\right)
}^{2}+\left\Vert f\right\Vert _{H^{4}\left( \Omega \right) }^{2}+\left\Vert
F\right\Vert _{H^{4}\left( \Omega \right) }^{2}\right) \geq
\end{equation*}%
\begin{equation}
\geq \dint\limits_{Q_{T}}\left(
w_{t}^{2}+\dsum\limits_{i,j=1}^{n}w_{x_{i}x_{j}}^{2}+\left( \nabla w\right)
^{2}+w^{2}\right) \phi ^{2}dxdt,\text{ }\forall \lambda \geq \lambda _{2}.
\label{6.9}
\end{equation}%
By (\ref{3.3}), (\ref{3.4}) and (\ref{3.22})%
\begin{equation*}
\phi ^{2}\left( x\right) \geq \exp \left( 2\lambda \left( \frac{3}{4}\right)
^{-\mu _{0}}\right) ,\text{ }x\in \Omega .
\end{equation*}%
Hence, 
\begin{equation*}
\dint\limits_{Q_{T}}\left(
w_{t}^{2}+\dsum\limits_{i,j=1}^{n}w_{x_{i}x_{j}}^{2}+\left( \nabla w\right)
^{2}+w^{2}\right) \phi ^{2}dxdt\geq \exp \left( 2\lambda \left( \frac{3}{4}%
\right) ^{-\mu _{0}}\right) \left\Vert w\right\Vert _{H^{2,1}\left(
Q_{T}\right) }^{2}.
\end{equation*}%
Substituting this in (\ref{6.9}), dividing the resulting inequality by $\exp
\left( 2\lambda \left( 3/4\right) ^{-\mu _{0}}\right) $ and setting then $%
\lambda =\lambda _{2},$ we obtain the following analog of (\ref{5.28}):%
\begin{equation}
\left\Vert w\right\Vert _{H^{2,1}\left( Q_{T}\right) }^{2}\leq C_{2}\left(
\left\Vert p_{t}\right\Vert _{H^{2,0}\left( S_{T}\right) }^{2}+\left\Vert
q_{t}\right\Vert _{H^{1,0}\left( S_{T}\right) }^{2}+\left\Vert f\right\Vert
_{H^{4}\left( \Omega \right) }^{2}+\left\Vert F\right\Vert _{H^{4}\left(
\Omega \right) }^{2}\right) .  \label{6.10}
\end{equation}

To finish the proof, we proceed similarly with the end of the proof of
Theorem 2. More precisely, we return from $w$ to $u$ via (\ref{5.1}) and (%
\ref{5.16}). Next, using (\ref{1.16}), (\ref{5.12}) and (\ref{6.10}), we
obtain the target estimates (\ref{3.25}) and (\ref{3.26}). \ $\square $

\section{Appendix: Proof of Theorem 1}

\label{sec:8}

This proof is inevitably space consuming, so as proofs of all pointwise
Carleman estimates. On the other hand, as it was pointed out in subsection
4.2, short proofs via symbols of operators would not deliver us boundary
terms (\ref{3.16}) and (\ref{3.160}), which we need for proofs of Theorems
2-4.

In this section $\left( x,t\right) \in G_{T}$ and $C=C\left( G,\nu ,A\right)
>0$ denotes different numbers depending only on the domain $G$ and the
numbers $\nu $ and $A$. In the course of the proof we do not fix the
parameter $\mu $, assuming only that $\mu \geq \mu _{0},$ where the number $%
\mu _{0}=\mu _{0}\left( G,\nu ,A\right) \geq 1$ is sufficiently large and
depends only on listed parameters. We set $\mu =\mu _{0}$ only in
sub-subsection 8.8.2.

Introduce a new function $w$, 
\begin{equation}
w=u\phi .  \label{4.0}
\end{equation}%
By (\ref{4.0}) $u=w\phi ^{-1}=w\exp \left( -\lambda \varphi ^{-\mu }\right)
. $ Using (\ref{3.3}) and (\ref{3.4}), express derivatives of the function $%
u $ via derivatives of the function $w$,%
\begin{equation}
\left. 
\begin{array}{c}
u_{t}=w_{t}\phi ^{-1}, \\ 
u_{i}=\left( w_{i}+\lambda \mu \varphi ^{-\mu -1}\varphi _{i}w\right) \phi
^{-1}, \\ 
u_{ij}=\left[ w_{ij}+\lambda \mu \varphi ^{-\mu -1}\left( \varphi
_{j}w_{i}+\varphi _{i}w_{j}\right) \right] \phi ^{-1}+ \\ 
+\lambda ^{2}\mu ^{2}\varphi ^{-2\mu -2}\left( \varphi _{i}\varphi
_{j}\left( 1-\lambda ^{-1}\left( 1+\mu ^{-1}\right) \varphi ^{\mu }\right)
+\left( \lambda \mu \right) ^{-1}\varphi ^{\mu +1}\varphi _{ij}\right) w\phi
^{-1}.%
\end{array}%
\right.  \label{4.1}
\end{equation}%
By (\ref{1.5}) and (\ref{4.1})%
\begin{equation}
\left. 
\begin{array}{c}
\left( u_{t}-L_{0}u\right) ^{2}\varphi ^{\mu +2}\phi ^{2}= \\ 
=\left[ 
\begin{array}{c}
w_{t}-\dsum\limits_{i,j=1}^{n}a^{ij}\left( x\right) w_{ij}-\lambda \mu
\varphi ^{-\mu -1}\dsum\limits_{i,j=1}^{n}a^{ij}\left( x\right) \left(
\varphi _{j}w_{i}+\varphi _{i}w_{j}\right) - \\ 
-\lambda ^{2}\mu ^{2}\varphi ^{-2\mu -2}\dsum\limits_{i,j=1}^{n}a^{ij}\left(
x\right) \times \\ 
\times \left[ \varphi _{i}\varphi _{j}\left( 1-\lambda ^{-1}\left( 1+\mu
^{-1}\right) \varphi ^{\mu }\right) +\left( \lambda \mu \right) ^{-1}\varphi
_{ij}\varphi ^{\mu +1}\right] w%
\end{array}%
\right] ^{2}\varphi ^{\mu +2}.%
\end{array}%
\right.  \label{4.2}
\end{equation}%
Denote%
\begin{equation}
\left. 
\begin{array}{c}
s_{1}=w_{t}, \\ 
s_{2}=-\dsum\limits_{i,j=1}^{n}a^{ij}\left( x\right) w_{ij}, \\ 
s_{3}=-\lambda \mu \varphi ^{-\mu -1}\dsum\limits_{i,j=1}^{n}a^{ij}\left(
x\right) \left( \varphi _{j}w_{i}+\varphi _{i}w_{j}\right) , \\ 
s_{4}=-\lambda ^{2}\mu ^{2}\varphi ^{-2\mu
-2}\dsum\limits_{i,j=1}^{n}a^{ij}\left( x\right) \times \\ 
\times \left[ \varphi _{i}\varphi _{j}\left( 1-\lambda ^{-1}\left( 1+\mu
^{-1}\right) \varphi ^{\mu }\right) +\left( \lambda \mu \right) ^{-1}\varphi
^{\mu +1}\varphi _{ij}\right] w.%
\end{array}%
\right.  \label{4.3}
\end{equation}%
By (\ref{4.2}) and (\ref{4.3})%
\begin{equation}
\left. 
\begin{array}{c}
\left( u_{t}-L_{0}u\right) ^{2}\varphi ^{\mu +2}\phi ^{2}=\left[ \left(
s_{1}+s_{3}\right) +\left( s_{2}+s_{4}\right) \right] ^{2}\varphi ^{\mu
+2}\geq \\ 
\geq \left[ \left( s_{1}+s_{3}\right) ^{2}+2\left( s_{1}+s_{3}\right) \left(
s_{2}+s_{4}\right) \right] \varphi ^{\mu +2}= \\ 
=\left( s_{1}^{2}+s_{3}^{2}+2s_{1}s_{2}+2s_{1}s_{3}\right) \varphi ^{\mu
+2}+2s_{2}s_{3}\varphi ^{\mu +2}+2s_{3}s_{4}\varphi ^{\mu
+2}+2s_{1}s_{4}\varphi ^{\mu +2}.%
\end{array}%
\right.  \label{4.4}
\end{equation}%
We estimate from the below all terms in the last line of (\ref{4.4})
one-by-one.

\subsection{Estimate from the below the term $2s_{1}s_{2}\protect\varphi ^{%
\protect\mu +2}$ in (\protect\ref{4.4})}

\label{sec:8.1}

By (\ref{1.3}) and (\ref{4.3})%
\begin{equation*}
2s_{1}s_{2}\varphi ^{\mu +2}=-\dsum\limits_{i,j=1}^{n}a^{ij}\left(
w_{ij}w_{t}+w_{ji}w_{t}\right) \varphi ^{\mu +2}=
\end{equation*}%
\begin{equation*}
=\dsum\limits_{i,j=1}^{n}\left[ \left( -a^{ij}w_{i}w_{t}\varphi ^{\mu
+2}\right) _{j}+\left( -a^{ij}w_{j}w_{t}\varphi ^{\mu +2}\right) _{i}\right]
+\dsum\limits_{i,j=1}^{n}a^{ij}\varphi ^{\mu +2}\left(
w_{i}w_{tj}+w_{j}w_{ti}\right) +
\end{equation*}%
\begin{equation*}
+\dsum\limits_{i,j=1}^{n}\left( a_{j}^{ij}w_{i}+a_{i}^{ij}w_{j}\right)
w_{t}\varphi ^{\mu +2}+\left( \mu +2\right) \varphi ^{\mu
+1}w_{t}\dsum\limits_{i,j=1}^{n}a^{ij}\left( \varphi _{j}w_{i}+\varphi
_{i}w_{j}\right) =
\end{equation*}%
\begin{equation*}
=\dsum\limits_{i,j=1}^{n}\left[ \left( -a^{ij}w_{i}w_{t}\varphi ^{\mu
+2}\right) _{j}+\left( -a^{ij}w_{j}w_{t}\varphi ^{\mu +2}\right) _{i}\right]
+\partial _{t}\left( \dsum\limits_{i,j=1}^{n}a^{ij}\varphi ^{\mu
+2}w_{i}w_{j}\right) +
\end{equation*}%
\begin{equation*}
+\left( \mu +2\right) \varphi ^{\mu +1}s_{1}\left[ \dsum%
\limits_{i,j=1}^{n}a^{ij}\left( \varphi _{j}w_{i}+\varphi _{i}w_{j}\right) +%
\frac{\varphi }{\left( \mu +2\right) }\dsum\limits_{i,j=1}^{n}\left(
a_{j}^{ij}w_{i}+a_{i}^{ij}w_{j}\right) \right] .
\end{equation*}%
Thus, 
\begin{equation*}
2s_{1}s_{2}\varphi ^{\mu +2}=
\end{equation*}%
\begin{equation*}
=\left( \mu +2\right) \varphi ^{\mu +1}s_{1}\left[ \dsum%
\limits_{i,j=1}^{n}a^{ij}\left( \varphi _{j}w_{i}+\varphi _{i}w_{j}\right) +%
\frac{\varphi }{\left( \mu +2\right) }\dsum\limits_{i,j=1}^{n}\left(
a_{j}^{ij}w_{i}+a_{i}^{ij}w_{j}\right) \right] +
\end{equation*}%
\begin{equation}
+\partial _{t}V_{1}+\func{div}U_{1},  \label{4.5}
\end{equation}%
\begin{equation}
\partial _{t}V_{1}=\partial _{t}\left( \dsum\limits_{i,j=1}^{n}a^{ij}\varphi
^{\mu +2}\left( u_{i}-\lambda \mu \varphi _{i}\varphi ^{-\mu -1}u\right)
\left( u_{j}-\lambda \mu \varphi _{j}\varphi ^{-\mu -1}u\right) \phi
^{2}\right) .  \label{4.7}
\end{equation}%
\begin{equation}
\func{div}U_{1}=\dsum\limits_{i,j=1}^{n}\left[ \left(
-a^{ij}w_{i}w_{t}\varphi ^{\mu +2}\right) _{j}+\left(
-a^{ij}w_{j}w_{t}\varphi ^{\mu +2}\right) _{i}\right] ,  \label{4.6}
\end{equation}%
\begin{equation}
w_{i}=\left( u_{i}-\lambda \mu \varphi _{i}\varphi ^{-\mu -1}u\right) \phi
,i=1,...,n.  \label{4.60}
\end{equation}

\subsection{Estimate from the below the term $\left(
s_{1}^{2}+s_{3}^{2}+2s_{1}s_{2}+2s_{1}s_{3}\right) \protect\varphi ^{\protect%
\mu +2}$ in (\protect\ref{4.4})}

\label{sec:8.2}

Using (\ref{4.3}), (\ref{4.5})-(\ref{4.7}) and Cauchy-Schwarz inequality, we
obtain 
\begin{equation*}
\left( s_{1}^{2}+s_{3}^{2}+2s_{1}s_{2}+2s_{1}s_{3}\right) \varphi ^{\mu
+2}=\left( s_{1}^{2}+s_{3}^{2}\right) \varphi ^{\mu +2}+
\end{equation*}%
\begin{equation*}
+2\left( \mu +2\right) \varphi ^{\mu +2}s_{1}\times
\end{equation*}%
\begin{equation*}
\times \left[ 
\begin{array}{c}
\left( 1/2\right) \dsum\limits_{i,j=1}^{n}a^{ij}\left( \varphi
_{j}w_{i}+\varphi _{i}w_{j}\right) \varphi ^{-1}+\left( 1/2\right)
\dsum\limits_{i,j=1}^{n}\left( a_{j}^{ij}w_{i}+a_{i}^{ij}w_{j}\right)
/\left( \mu +2\right) \\ 
+s_{3}/\left( \mu +2\right)%
\end{array}%
\right] +
\end{equation*}%
\begin{equation*}
+\partial _{t}V_{1}+\func{div}U_{1}\geq \left( s_{1}^{2}+s_{3}^{2}\right)
\varphi ^{\mu +2}-s_{1}^{2}\varphi ^{\mu +2}-\left( \mu +2\right)
^{2}\varphi ^{\mu +2}\times
\end{equation*}%
\begin{equation*}
\times \left[ \frac{1}{2}\dsum\limits_{i,j=1}^{n}a^{ij}\left( \varphi
_{j}w_{i}+\varphi _{i}w_{j}\right) \varphi ^{-1}+\frac{1}{2\left( \mu
+2\right) }\dsum\limits_{i,j=1}^{n}\left(
a_{j}^{ij}w_{i}+a_{i}^{ij}w_{j}\right) +\frac{s_{3}}{\left( \mu +2\right) }%
\right] ^{2}+
\end{equation*}%
\begin{equation}
+\partial _{t}V_{1}+\func{div}U_{1}.  \label{4}
\end{equation}%
Next, by (\ref{4.3})%
\begin{equation*}
-\left( \mu +2\right) ^{2}\varphi ^{\mu +2}\times
\end{equation*}%
\begin{equation*}
\times \left[ \frac{1}{2}\dsum\limits_{i,j=1}^{n}a^{ij}\left( \varphi
_{j}w_{i}+\varphi _{i}w_{j}\right) \varphi ^{-1}+\frac{1}{2\left( \mu
+2\right) }\dsum\limits_{i,j=1}^{n}\left(
a_{j}^{ij}w_{i}+a_{i}^{ij}w_{j}\right) +\frac{s_{3}}{\left( \mu +2\right) }%
\right] ^{2}\geq
\end{equation*}%
\begin{equation*}
\geq -s_{3}^{2}\varphi ^{\mu +2}+\lambda \left( \mu +2\right) \mu \left[
\dsum\limits_{i,j=1}^{n}a^{ij}\left( x\right) \left( \varphi
_{j}w_{i}+\varphi _{i}w_{j}\right) \right] ^{2}+
\end{equation*}%
\begin{equation}
+\lambda \mu \varphi \left( \dsum\limits_{i,j=1}^{n}a^{ij}\left( x\right)
\left( \varphi _{j}w_{i}+\varphi _{i}w_{j}\right) \right) \left(
\dsum\limits_{i,j=1}^{n}\left( a_{j}^{ij}w_{i}+a_{i}^{ij}w_{j}\right)
\right) -  \label{4.8}
\end{equation}%
\begin{equation*}
-\left( \mu +2\right) ^{2}\varphi ^{\mu +2}\left[ \frac{1}{2}%
\dsum\limits_{i,j=1}^{n}a^{ij}\left( \varphi _{j}w_{i}+\varphi
_{i}w_{j}\right) \varphi ^{-1}+\frac{1}{2\left( \mu +2\right) }%
\dsum\limits_{i,j=1}^{n}\left( a_{j}^{ij}w_{i}+a_{i}^{ij}w_{j}\right) \right]
^{2}.
\end{equation*}%
Combining (\ref{4}) with (\ref{4.8}) and dropping the non-negative second
term in the third line of (\ref{4.8}), we obtain 
\begin{equation*}
\left( s_{1}^{2}+s_{3}^{2}+2s_{1}s_{2}+2s_{1}s_{3}\right) \varphi ^{\mu
+2}\geq
\end{equation*}%
\begin{equation*}
\geq -\left( \mu +2\right) ^{2}\varphi ^{\mu +2}\left[ \frac{1}{2}%
\dsum\limits_{i,j=1}^{n}a^{ij}\left( \varphi _{j}w_{i}+\varphi
_{i}w_{j}\right) \varphi ^{-1}+\frac{1}{2\left( \mu +2\right) }%
\dsum\limits_{i,j=1}^{n}\left( a_{j}^{ij}w_{i}+a_{i}^{ij}w_{j}\right) \right]
^{2}+
\end{equation*}%
\begin{equation*}
+\lambda \mu \varphi \left( \dsum\limits_{i,j=1}^{n}a^{ij}\left( x\right)
\left( \varphi _{j}w_{i}+\varphi _{i}w_{j}\right) \right) \left(
\dsum\limits_{i,j=1}^{n}\left( a_{j}^{ij}w_{i}+a_{i}^{ij}w_{j}\right)
\right) +
\end{equation*}%
\begin{equation}
+\partial _{t}V_{1}+\func{div}U_{1}.  \label{5}
\end{equation}%
Since by (\ref{3.3}) and (\ref{3.5}) $\varphi \in \left[ 1/4,3/4\right] $ in 
$G_{T},$ then $\left( \mu +2\right) ^{2}\varphi ^{\mu +2}<1$ for $\mu \geq
\mu _{0}=\mu _{0}\left( G\right) >0$. Hence, the term in the second line of (%
\ref{5}) can be estimated as:%
\begin{equation*}
-\left( \mu +2\right) ^{2}\varphi ^{\mu +2}\left[ \frac{1}{2}%
\dsum\limits_{i,j=1}^{n}a^{ij}\left( \varphi _{j}w_{i}+\varphi
_{i}w_{j}\right) \varphi ^{-1}+\frac{1}{\left( \mu +2\right) }%
\dsum\limits_{i,j=1}^{n}\left( a_{j}^{ij}w_{i}+a_{i}^{ij}w_{j}\right) \right]
^{2}\geq
\end{equation*}%
\begin{equation}
\geq -C\left( \nabla w\right) ^{2}.  \label{4.90}
\end{equation}%
Next, (\ref{3.4}) and (\ref{4.0}), we obtain from (\ref{4.90}) 
\begin{equation*}
-\left( \mu +2\right) ^{2}\varphi ^{\mu +2}\left[ \frac{1}{2}%
\dsum\limits_{i,j=1}^{n}a^{ij}\left( \varphi _{j}w_{i}+\varphi
_{i}w_{j}\right) \varphi ^{-1}+\frac{1}{\left( \mu +2\right) }%
\dsum\limits_{i,j=1}^{n}\left( a_{j}^{ij}w_{i}+a_{i}^{ij}w_{j}\right) \right]
^{2}\geq
\end{equation*}%
\begin{equation}
\geq -C\left( \nabla u\right) ^{2}\phi ^{2}-C\lambda ^{2}\mu ^{2}\varphi
^{-2\mu -2}u^{2}\phi ^{2}.  \label{4.9}
\end{equation}%
The term in the third line of (\ref{5}) can be estimated as: 
\begin{equation*}
\lambda \mu \varphi \dsum\limits_{i,j=1}^{n}a^{ij}\left( x\right) \left(
\varphi _{j}w_{i}+\varphi _{i}w_{j}\right) \cdot
\dsum\limits_{i,j=1}^{n}\left( a_{j}^{ij}w_{i}+a_{i}^{ij}w_{j}\right) \geq
-C\lambda \mu \left( \nabla w\right) ^{2}\geq
\end{equation*}%
\begin{equation}
\geq -C\lambda \mu \left( \nabla u\right) ^{2}\phi ^{2}-C\lambda ^{3}\mu
^{3}\varphi ^{-2\mu -2}u^{2}\phi ^{2}.  \label{4.10}
\end{equation}%
Thus, (\ref{5})-(\ref{4.10}) imply:%
\begin{equation*}
\left( s_{1}^{2}+s_{3}^{2}+2s_{1}s_{2}+2s_{1}s_{3}\right) \varphi ^{\mu
+2}\geq -C\lambda \mu \left( \nabla u\right) ^{2}\phi ^{2}-C\lambda ^{3}\mu
^{3}\varphi ^{-2\mu -2}u^{2}\phi ^{2}+
\end{equation*}%
\begin{equation}
+\partial _{t}V_{1}+\func{div}U_{1},  \label{4.11}
\end{equation}%
where $V_{1}$ and $\func{div}U_{1}$ are given in (\ref{4.7}) and (\ref{4.6})
respectively.

\subsection{Estimate from the below the term $2s_{2}s_{3}\protect\varphi ^{%
\protect\mu +2}$ in (\protect\ref{4.4})}

\label{sec:8.3}

Using (\ref{4.3}), we obtain 
\begin{equation*}
2s_{2}s_{3}\varphi ^{\mu +2}=\lambda \mu \varphi
\dsum\limits_{i,j,k.s=1}^{n}a^{ij}\left( x\right) a^{ks}\left( x\right)
w_{ij}\left( \varphi _{s}w_{k}+\varphi _{k}w_{s}\right) .
\end{equation*}%
Consider the term%
\begin{equation*}
\lambda \mu \varphi a^{ij}\left( x\right) a^{ks}\left( x\right) w_{ij}\left(
\varphi _{s}w_{k}\right) +\lambda \mu \varphi a^{ji}\left( x\right)
a^{ks}\left( x\right) w_{ji}\left( \varphi _{s}w_{k}\right) =
\end{equation*}%
\begin{equation*}
=\lambda \mu a^{ij}\left( x\right) a^{ks}\left( x\right) \varphi \varphi
_{s}\left( w_{ij}w_{k}+w_{ji}w_{k}\right) =
\end{equation*}%
\begin{equation*}
=\lambda \mu a^{ij}\left( x\right) a^{ks}\left( x\right) \left[ \left(
w_{i}w_{k}\right) _{j}+\left( w_{j}w_{k}\right) _{i}-w_{i}w_{kj}-w_{j}w_{ki}%
\right] =
\end{equation*}%
\begin{equation*}
=\lambda \mu a^{ij}\left( x\right) a^{ks}\left( x\right) \left[ \left(
w_{i}w_{k}\right) _{j}+\left( w_{j}w_{k}\right) _{i}+\left(
-w_{i}w_{j}\right) _{k}\right] =
\end{equation*}%
\begin{equation*}
=\left( \lambda \mu a^{ij}\left( x\right) a^{ks}\left( x\right)
w_{i}w_{k}\right) _{j}+\left( \lambda \mu a^{ij}\left( x\right) a^{ks}\left(
x\right) w_{j}w_{k}\right) _{i}+\left( -\lambda \mu a^{ij}\left( x\right)
a^{ks}\left( x\right) w_{i}w_{j}\right) _{k}-
\end{equation*}%
\begin{equation*}
-\lambda \mu \left[ \left( a^{ij}\left( x\right) a^{ks}\left( x\right)
\right) _{j}w_{i}w_{k}+\left( a^{ij}\left( x\right) a^{ks}\left( x\right)
\right) _{i}w_{j}w_{k}-\left( a^{ij}\left( x\right) a^{ks}\left( x\right)
\right) _{k}w_{i}w_{j}\right] .
\end{equation*}%
Hence, applying the backwards substitution (\ref{4.0}) and using (\ref{3.3})
and (\ref{3.4}), we obtain 
\begin{equation}
2s_{2}s_{3}\varphi ^{\mu +2}\geq -C\lambda \mu \left( \nabla u\right)
^{2}\phi ^{2}-C\lambda ^{3}\mu ^{3}u^{2}\phi ^{2}+\func{div}U_{2},
\label{4.12}
\end{equation}%
\begin{equation}
\func{div}U_{2}=\dsum\limits_{i,j,k,s=1}^{n}\left[ 
\begin{array}{c}
\left( \lambda \mu a^{ij}\left( x\right) a^{ks}\left( x\right)
w_{i}w_{k}\right) _{j}+\left( \lambda \mu a^{ij}\left( x\right) a^{ks}\left(
x\right) w_{j}w_{k}\right) _{i}+ \\ 
+\left( -\lambda \mu a^{ij}\left( x\right) a^{ks}\left( x\right)
w_{i}w_{j}\right) _{k}%
\end{array}%
\right] ,  \label{4.13}
\end{equation}%
where $w_{i}$ are as in (\ref{4.60}) and similarly for $w_{j}$ and $w_{k}$.

\subsection{Estimate from the below the term $2s_{3}s_{4}\protect\varphi ^{%
\protect\mu +2}$ in (\protect\ref{4.4})}

\label{sec:8.4}

Using (\ref{4.3}), we obtain%
\begin{equation*}
2s_{3}s_{4}\varphi ^{\mu +2}=2\lambda ^{3}\mu ^{3}\varphi ^{-2\mu -1}\times
\end{equation*}%
\begin{equation}
\times \dsum\limits_{i,j,k,s=1}^{n}a^{ij}\left( x\right) a^{ks}\left(
x\right) \left[ \varphi _{k}\varphi _{s}\left( \varphi _{j}w_{i}+\varphi
_{i}w_{j}\right) w\right] \left( 1-\lambda ^{-1}\left( 1+\mu ^{-1}\right)
\varphi ^{\mu }\right) +  \label{4.130}
\end{equation}%
\begin{equation*}
+2\lambda ^{2}\mu ^{2}\varphi ^{-\mu
}\dsum\limits_{i,j,k,s=1}^{n}a^{ij}\left( x\right) a^{ks}\left( x\right)
\varphi _{ks}\left( \varphi _{j}w_{i}+\varphi _{i}w_{j}\right) w.
\end{equation*}%
We have:%
\begin{equation*}
2\varphi ^{-2\mu -1}a^{ij}\left( x\right) a^{ks}\left( x\right) \varphi
_{k}\varphi _{s}\left( \varphi _{j}w_{i}+\varphi _{i}w_{j}\right) w\left[
\left( 1-\lambda ^{-1}\left( 1+\mu ^{-1}\right) \varphi ^{\mu }\right) %
\right] =
\end{equation*}%
\begin{equation*}
=\left[ \varphi ^{-2\mu -1}a^{ij}\left( x\right) a^{ks}\left( x\right)
\varphi _{k}\varphi _{s}\varphi _{j}\left( \left( 1-\lambda ^{-1}\left(
1+\mu ^{-1}\right) \varphi ^{\mu }\right) \right) w^{2}\right] _{i}+
\end{equation*}%
\begin{equation*}
+\left[ \varphi ^{-2\mu -1}a^{ij}\left( x\right) a^{ks}\left( x\right)
\varphi _{k}\varphi _{s}\varphi _{i}\left( \left( 1-\lambda ^{-1}\left(
1+\mu ^{-1}\right) \varphi ^{\mu }\right) \right) w^{2}\right] _{j}+
\end{equation*}%
\begin{equation*}
+2\left( 2\mu +1\right) \varphi ^{-2\mu -2}\left[ a^{ij}\left( x\right)
a^{ks}\left( x\right) \varphi _{k}\varphi _{s}\varphi _{j}\varphi _{i}\right]
w^{2}+B\left( x,\mu \right) \varphi ^{-2\mu -1}w^{2},
\end{equation*}%
where $\left\vert B\left( x,\mu \right) \right\vert \leq C$ for all $\mu
\geq \mu _{0}$ and for all $x\in \overline{G},$ and also $B\left( x,\mu
\right) $ is independent on $w$. Hence,%
\begin{equation*}
\dsum\limits_{i,j,k,s=1}^{n}a^{ij}\left( x\right) a^{ks}\left( x\right) 
\left[ \varphi _{k}\varphi _{s}\left( \varphi _{j}w_{i}+\varphi
_{i}w_{j}\right) w\right] \left( 1-\lambda ^{-1}\left( 1+\mu ^{-1}\right)
\varphi ^{\mu }\right) \geq
\end{equation*}
\begin{equation}
\geq C\lambda ^{3}\mu ^{4}\varphi ^{-2\mu -2}\phi ^{2}u^{2}+\func{div}U_{3},
\label{4.15}
\end{equation}%
\begin{equation*}
\func{div}U_{3}=
\end{equation*}%
\begin{equation}
=\dsum\limits_{i,j,k,s=1}^{n}\left[ \varphi ^{-2\mu -1}a^{ij}\left( x\right)
a^{ks}\left( x\right) \varphi _{k}\varphi _{s}\varphi _{j}\left( \left(
1-\lambda ^{-1}\left( 1+\mu ^{-1}\right) \varphi ^{\mu }\right) \right) \phi
^{2}u^{2}\right] _{i}+  \label{4.16}
\end{equation}%
\begin{equation*}
+\dsum\limits_{i,j,k,s=1}^{n}\left[ \varphi ^{-2\mu -1}a^{ij}\left( x\right)
a^{ks}\left( x\right) \varphi _{k}\varphi _{s}\varphi _{i}\left( \left(
1-\lambda ^{-1}\left( 1+\mu ^{-1}\right) \varphi ^{\mu }\right) \right) \phi
^{2}u^{2}\right] _{j}.
\end{equation*}

We now estimate from the below the term in the third line of (\ref{4.130}).
Using Cauchy-Schwarz inequality and (\ref{4.60}), we obtain 
\begin{equation*}
2\lambda ^{2}\mu ^{2}\varphi ^{-\mu
}\dsum\limits_{i,j,k,s=1}^{n}a^{ij}\left( x\right) a^{ks}\left( x\right)
\varphi _{ks}\left( \varphi _{j}w_{i}+\varphi _{i}w_{j}\right) w\geq
\end{equation*}%
\begin{equation*}
\geq -C\lambda \mu \left\vert \nabla w\right\vert ^{2}-C\lambda ^{3}\mu
^{3}\varphi ^{-2\mu -2}w^{2}\geq -C\lambda \mu \left\vert \nabla
u\right\vert ^{2}\phi ^{2}-C\lambda ^{3}\mu ^{3}\varphi ^{-2\mu -2}u^{2}\phi
^{2}.
\end{equation*}%
Combining this with (\ref{4.130}) and (\ref{4.15}), we obtain%
\begin{equation}
2s_{3}s_{4}\varphi ^{\mu +2}\geq -C\lambda \mu \left\vert \nabla
u\right\vert ^{2}\phi ^{2}+C\lambda ^{3}\mu ^{4}\varphi ^{-2\mu -2}\phi
^{2}u^{2}+\func{div}U_{3}.  \label{4.160}
\end{equation}

\subsection{Estimate from the below the term $2s_{1}s_{4}\protect\varphi ^{%
\protect\mu +2}$ in (\protect\ref{4.4})}

\label{sec:8.5}

Using (\ref{4.0}) and (\ref{4.3}), we obtain%
\begin{equation*}
2s_{1}s_{4}\varphi ^{\mu +2}=-2\lambda ^{2}\mu ^{2}\varphi ^{-\mu }\times
\end{equation*}%
\begin{equation*}
\times \dsum\limits_{i,j=1}^{n}a^{ij}\left( x\right) \left[ \varphi
_{i}\varphi _{j}\left( 1-\lambda ^{-1}\left( 1+\mu ^{-1}\right) \varphi
^{\mu }\right) +\left( \lambda \mu \right) ^{-1}\varphi ^{\mu +1}\varphi
_{ij}\right] ww_{t}=
\end{equation*}%
\begin{equation*}
=\partial _{t}\left( -\lambda ^{2}\mu ^{2}\varphi ^{-\mu
}\dsum\limits_{i,j=1}^{n}a^{ij}\left( x\right) \left[ \varphi _{i}\varphi
_{j}\left( 1-\lambda ^{-1}\left( 1+\mu ^{-1}\right) \varphi ^{\mu }\right)
+\left( \lambda \mu \right) ^{-1}\varphi ^{\mu +1}\varphi _{ij}\right]
w^{2}\right) =
\end{equation*}%
\begin{equation}
=\partial _{t}V_{2}.  \label{4.17}
\end{equation}

\subsection{Sum up estimates (\protect\ref{4.11})-(\protect\ref{4.13}), (%
\protect\ref{4.160}) and (\protect\ref{4.17}) and use (\protect\ref{4.4}), (%
\protect\ref{4.5}), (\protect\ref{4.7}), (\protect\ref{4.6}) and (\protect
\ref{4.16})}

\label{sec:8.6}

Recall that $\mu \geq \mu _{0}$ and $\mu _{0}$ is sufficiently large. Since
we have the term $C\lambda ^{3}\mu ^{4}\varphi ^{-2\mu -2}\phi ^{2}u^{2}$ in
(\ref{4.160}) and since $C\lambda ^{3}\mu ^{4}\varphi ^{-2\mu -2}>>C\lambda
^{3}\mu ^{3}\varphi ^{-2\mu -2}$ for $\mu \geq \mu _{0}$ (see (\ref{4.12})),
then we obtain%
\begin{equation*}
\left( u_{t}-L_{0}u\right) ^{2}\varphi ^{\mu +2}\phi ^{2}\geq -C\lambda \mu
\left( \nabla u\right) ^{2}\phi ^{2}+C\lambda ^{3}\mu ^{4}\varphi ^{-2\mu
-2}u^{2}\phi ^{2}+
\end{equation*}%
\begin{equation*}
+\func{div}\left( U_{1}+U_{2}+U_{3}\right) +
\end{equation*}%
\begin{equation}
+\partial _{t}\left( \dsum\limits_{i,j=1}^{n}a^{ij}\left( x\right) \varphi
^{\mu +2}\left( u_{i}-\lambda \mu \varphi _{i}\varphi ^{-\mu -1}u\right)
\left( u_{j}-\lambda \mu \varphi _{j}\varphi ^{-\mu -1}u\right) \phi
^{2}\right) +  \label{4.18}
\end{equation}%
\begin{equation*}
+\partial _{t}\left( -\lambda ^{2}\mu ^{2}\varphi ^{-\mu
}\dsum\limits_{i,j=1}^{n}a^{ij}\left( x\right) \left[ \varphi _{i}\varphi
_{j}\left( 1-\lambda ^{-1}\left( 1+\mu ^{-1}\right) \varphi ^{\mu }\right)
+\left( \lambda \mu \right) ^{-1}\varphi ^{\mu +1}\varphi _{ij}\right]
u^{2}\phi ^{2}\right) ,
\end{equation*}%
where vector functions $U_{1},U_{2}$ and $U_{3}$ are given in (\ref{4.6}), (%
\ref{4.13}) and (\ref{4.16}), also see (\ref{4.60}). We need to balance the
negative term $-C\lambda \mu \phi ^{2}\left( \nabla u\right) ^{2}$ in the
first line of (\ref{4.18}). To do this, consider 
\begin{equation*}
\left( u_{t}-L_{0}u\right) u\phi ^{2}=\partial _{t}\left( \frac{u^{2}}{2}%
\phi ^{2}\right) +\dsum\limits_{i,j=1}^{n}\left( -a^{ij}\left( x\right)
u_{i}u\phi ^{2}\right) _{j}+
\end{equation*}%
\begin{equation*}
+\dsum\limits_{i,j=1}^{n}a^{ij}\left( x\right) u_{i}u_{j}\phi ^{2}-2\lambda
\mu \varphi ^{-\mu -1}\dsum\limits_{i,j=1}^{n}a^{ij}\left( x\right) \varphi
_{j}u_{i}u\phi ^{2}+\dsum\limits_{i,j=1}^{n}a_{j}^{ij}\left( x\right)
u_{i}u\phi ^{2}\geq
\end{equation*}%
\begin{equation*}
\geq C\left( \nabla u\right) ^{2}\phi ^{2}-C\lambda ^{2}\mu ^{2}\varphi
^{-2\mu -2}\phi ^{2}u^{2}+\partial _{t}\left( \frac{u^{2}}{2}\phi
^{2}\right) +\dsum\limits_{i,j=1}^{n}\left( -a^{ij}\left( x\right)
u_{i}u\phi ^{2}\right) _{j}.
\end{equation*}%
Thus, 
\begin{equation*}
\left( u_{t}-L_{0}u\right) u\phi ^{2}\geq C\left( \nabla u\right) ^{2}\phi
^{2}-C\lambda ^{2}\mu ^{2}\varphi ^{-2\mu -2}\phi ^{2}u^{2}+
\end{equation*}%
\begin{equation}
+\func{div}U_{4}+\partial _{t}\left( \frac{u^{2}}{2}\phi ^{2}\right) ,
\label{4.19}
\end{equation}%
\begin{equation}
\func{div}U_{4}=\dsum\limits_{i,j=1}^{n}\left( -a^{ij}\left( x\right)
u_{i}u\phi ^{2}\right) _{j}.  \label{4.20}
\end{equation}

\subsection{Estimate $\left( u_{t}-L_{0}u\right) ^{2}\protect\phi ^{2}$ from
the below}

\label{sec:8.7}

Multiply (\ref{4.19}) by $2\lambda \mu $ and sum up with (\ref{4.18}). Since 
$\lambda ^{3}\mu ^{4}\varphi ^{-2\mu -2}>>\lambda ^{3}\mu ^{3}\varphi
^{-2\mu -2}$ for all $\mu \geq \mu _{0},$ we obtain 
\begin{equation*}
\left( u_{t}-L_{0}u\right) ^{2}\phi ^{2}+2\lambda \mu \left(
u_{t}-L_{0}u\right) u\phi ^{2}\geq
\end{equation*}%
\begin{equation*}
\geq C\lambda \mu \phi ^{2}\left( \nabla u\right) ^{2}+C\lambda ^{3}\mu
^{4}\varphi ^{-2\mu -2}\phi ^{2}u^{2}+
\end{equation*}%
\begin{equation*}
+\func{div}\left( U_{1}+U_{2}+U_{3}+U_{4}\right) +
\end{equation*}%
\begin{equation}
+\partial _{t}\left( \dsum\limits_{i,j=1}^{n}a^{ij}\left( x\right) \varphi
^{\mu +2}\left( u_{i}-\lambda \mu \varphi _{i}\varphi ^{-\mu -1}u\right)
\left( u_{j}-\lambda \mu \varphi _{j}\varphi ^{-\mu -1}u\right) \phi
^{2}\right) +  \label{4.21}
\end{equation}%
\begin{equation*}
+\partial _{t}\left( -\lambda ^{2}\mu ^{2}\varphi ^{-\mu
}\dsum\limits_{i,j=1}^{n}a^{ij}\left( x\right) \left[ \varphi _{i}\varphi
_{j}\left( 1-\lambda ^{-1}\left( 1+\mu ^{-1}\right) \varphi ^{\mu }\right)
+\left( \lambda \mu \right) ^{-1}\varphi ^{\mu +1}\varphi _{ij}\right]
u^{2}\phi ^{2}\right) +
\end{equation*}%
\begin{equation*}
+\partial _{t}\left( \lambda \mu \phi ^{2}u^{2}\right) ,
\end{equation*}%
where $U_{4}$ is defined in (\ref{4.20}). Next,%
\begin{equation*}
\left( u_{t}-L_{0}u\right) ^{2}\phi ^{2}+2\lambda \mu \left(
u_{t}-L_{0}u\right) u\phi ^{2}\leq 2\left( u_{t}-L_{0}u\right) ^{2}\phi
^{2}+\lambda ^{2}\mu ^{2}u^{2}\phi ^{2}.
\end{equation*}%
Comparing this with (\ref{4.21}), we obtain%
\begin{equation*}
\left( u_{t}-L_{0}u\right) ^{2}\phi ^{2}\geq C\lambda \mu \phi ^{2}\left(
\nabla u\right) ^{2}+C\lambda ^{3}\mu ^{4}\varphi ^{-2\mu -2}\phi ^{2}u^{2}+
\end{equation*}%
\begin{equation*}
+\func{div}\left( U_{1}/2+U_{2}/2+U_{3}/2+U_{4}/2\right) +
\end{equation*}%
\begin{equation}
+\partial _{t}\left( \frac{1}{2}\dsum\limits_{i,j=1}^{n}a^{ij}\left(
x\right) \varphi ^{\mu +2}\left( u_{i}-\lambda \mu \varphi _{i}\varphi
^{-\mu -1}u\right) \left( u_{j}-\lambda \mu \varphi _{j}\varphi ^{-\mu
-1}u\right) \phi ^{2}\right) +  \label{4.22}
\end{equation}%
\begin{equation*}
+\partial _{t}\left( -\frac{1}{2}\lambda ^{2}\mu ^{2}\varphi ^{-\mu
}\dsum\limits_{i,j=1}^{n}a^{ij}\left( x\right) \left[ \varphi _{i}\varphi
_{j}\left( 1-\lambda ^{-1}\left( 1+\mu ^{-1}\right) \varphi ^{\mu }\right)
+\left( \lambda \mu \right) ^{-1}\varphi ^{\mu +1}\varphi _{ij}\right]
u^{2}\phi ^{2}\right) +
\end{equation*}%
\begin{equation*}
+\partial _{t}\left( \frac{1}{2}\lambda \mu \phi ^{2}u^{2}\right) ,
\end{equation*}%
where vector functions $U_{1},U_{2},U_{3},U_{4}$ are given in (\ref{4.6}), (%
\ref{4.60}), (\ref{4.13}), (\ref{4.16}) and (\ref{4.20}). Estimate (\ref%
{4.22}) is the pointwise Carleman estimate, in which lower order derivatives
are estimated in the first line of (\ref{4.22}). We now need to incorporate
in (\ref{4.22}) an estimate of the second order $x-$derivatives and the
first $t-$derivative of the function $u$.

\subsection{Estimate the sum of $u_{ij}^{2}\protect\phi ^{2}$ and $u_{t}^{2}%
\protect\phi ^{2}$ from the below}

\label{sec:8.8}

We have%
\begin{equation}
\left( u_{t}-L_{0}u\right) ^{2}\phi ^{2}=u_{t}^{2}\phi ^{2}+\left(
L_{0}u\right) ^{2}\phi ^{2}-2u_{t}L_{0}u\phi ^{2}.  \label{4.23}
\end{equation}

\subsubsection{Estimate the term $u_{t}^{2}\protect\phi ^{2}-2u_{t}L_{0}u%
\protect\phi ^{2}$ from the below}

\label{sec:8.8.1}

We have 
\begin{equation*}
u_{t}^{2}\phi ^{2}-2u_{t}L_{0}u\phi ^{2}=u_{t}^{2}\phi
^{2}-2\dsum\limits_{i,j=1}^{n}a^{ij}\left( x\right) u_{t}u_{ij}\phi
^{2}=u_{t}^{2}\phi ^{2}+\dsum\limits_{i,j=1}^{n}\left( -2a^{ij}\left(
x\right) u_{t}u_{i}\phi ^{2}\right) _{j}+
\end{equation*}%
\begin{equation*}
+\dsum\limits_{i,j=1}^{n}a^{ij}\left( x\right) \left(
u_{jt}u_{i}+u_{it}u_{j}\right) \phi ^{2}+u_{t}\dsum\limits_{i,j=1}^{n}\left(
2a_{j}^{ij}\left( x\right) u_{i}\phi ^{2}-4\lambda \mu \varphi ^{-\mu
-1}\varphi _{j}u_{i}\right) \phi ^{2}\geq
\end{equation*}%
\begin{equation*}
\geq \frac{u_{t}^{2}}{2}\phi ^{2}-C\lambda ^{2}\mu ^{2}\varphi ^{-2\mu
-2}\phi ^{2}\left( \nabla u\right) ^{2}+
\end{equation*}%
\begin{equation*}
+\dsum\limits_{i,j=1}^{n}\left( -2a^{ij}\left( x\right) u_{t}u_{i}\phi
^{2}\right) _{j}+\partial _{t}\left( \dsum\limits_{i,j=1}^{n}a^{ij}\left(
x\right) u_{i}u_{j}\phi ^{2}\right) .
\end{equation*}%
Thus, 
\begin{equation*}
u_{t}^{2}\phi ^{2}-2u_{t}L_{0}u\phi ^{2}\geq
\end{equation*}%
\begin{equation}
\geq \frac{u_{t}^{2}}{2}\phi ^{2}-C\lambda ^{2}\mu ^{2}\varphi ^{-2\mu
-2}\phi ^{2}\left( \nabla u\right) ^{2}+  \label{4.24}
\end{equation}%
\begin{equation*}
+\func{div}U_{5}+\partial _{t}\left( \dsum\limits_{i,j=1}^{n}a^{ij}\left(
x\right) u_{i}u_{j}\phi ^{2}\right) ,
\end{equation*}%
\begin{equation}
\func{div}U_{5}=\dsum\limits_{i,j=1}^{n}\left( -2a^{ij}\left( x\right)
u_{t}u_{i}\phi ^{2}\right) _{j}.  \label{4.240}
\end{equation}

\subsubsection{Estimate the term $\left( L_{0}u\right) ^{2}\protect\phi ^{2}$
from the below}

\label{sec:8.8.2}

We have%
\begin{equation}
\left( L_{0}u\right) ^{2}\phi ^{2}=\dsum\limits_{i,j,k,s=1}^{n}a^{ij}\left(
x\right) a^{ks}\left( x\right) u_{ij}u_{ks}\phi ^{2}.  \label{4.25}
\end{equation}%
Next,%
\begin{equation}
\left. 
\begin{array}{c}
a^{ij}\left( x\right) a^{ks}\left( x\right) u_{ij}u_{ks}\phi ^{2}=\left(
a^{ij}\left( x\right) a^{ks}\left( x\right) u_{i}u_{ks}\phi ^{2}\right)
_{j}-a^{ij}\left( x\right) a^{ks}\left( x\right) u_{i}u_{ksj}\phi ^{2}+ \\ 
+2\lambda \mu \varphi _{j}\varphi ^{-\mu -1}a^{ij}\left( x\right)
a^{ks}\left( x\right) u_{i}u_{ks}\phi ^{2}-\left( a^{ij}\left( x\right)
a^{ks}\left( x\right) \right) _{j}u_{i}u_{ks}\phi ^{2}= \\ 
=a^{ij}\left( x\right) a^{ks}\left( x\right) u_{ik}u_{sj}\phi ^{2}- \\ 
-2\lambda \mu \varphi _{k}\varphi ^{-\mu -1}a^{ij}\left( x\right)
a^{ks}\left( x\right) u_{i}u_{sj}\phi ^{2}+\left( a^{ij}\left( x\right)
a^{ks}\left( x\right) \right) _{k}u_{i}u_{sj}\phi ^{2}+ \\ 
+2\lambda \mu \varphi _{j}\varphi ^{-\mu -1}a^{ij}\left( x\right)
a^{ks}\left( x\right) u_{i}u_{ks}\phi ^{2}-\left( a^{ij}\left( x\right)
a^{ks}\left( x\right) \right) _{j}u_{i}u_{ks}\phi ^{2}+ \\ 
+\left( a^{ij}\left( x\right) a^{ks}\left( x\right) u_{i}u_{ks}\phi
^{2}\right) _{j}+\left( -a^{ij}\left( x\right) a^{ks}\left( x\right)
u_{i}u_{sj}\phi ^{2}\right) _{k}.%
\end{array}%
\right.  \label{4.26}
\end{equation}

It was proven in \cite[Chapter 2, formula (6.12)]{Lad} that 
\begin{equation}
\dsum\limits_{i,j,k,s=1}^{n}a^{ij}\left( x\right) a^{ks}\left( x\right)
u_{ik}u_{sj}\phi ^{2}\geq \nu ^{2}\dsum\limits_{i,j=1}^{n}u_{ij}^{2}\phi
^{2},  \label{4.27}
\end{equation}%
where $\nu >0$ is the number in (\ref{1.4}). Hence, (\ref{4.25})-(\ref{4.27}%
) and Cauchy-Schwarz inequality imply%
\begin{equation}
\left( L_{0}u\right) ^{2}\phi ^{2}\geq
C\dsum\limits_{i,j=1}^{n}u_{ij}^{2}\phi ^{2}-C\lambda ^{2}\mu ^{2}\varphi
^{-2\mu -2}\phi ^{2}\left( \nabla u\right) ^{2}+\func{div}U_{6},
\label{4.28}
\end{equation}%
\begin{equation}
\func{div}U_{6}=\dsum\limits_{i,j,k,s=1}^{n}\left[ \left( a^{ij}\left(
x\right) a^{ks}\left( x\right) u_{i}u_{ks}\phi ^{2}\right) _{j}+\left(
-a^{ij}\left( x\right) a^{ks}\left( x\right) u_{i}u_{sj}\phi ^{2}\right) _{k}%
\right] .  \label{4.29}
\end{equation}%
Thus, using (\ref{4.23})-(\ref{4.240}) and (\ref{4.28}), we obtain 
\begin{equation*}
\left( u_{t}-L_{0}u\right) ^{2}\phi ^{2}\geq \frac{u_{t}^{2}}{2}\phi
^{2}+C\dsum\limits_{i,j=1}^{n}u_{ij}^{2}\phi ^{2}-C\lambda ^{2}\mu
^{2}\varphi ^{-2\mu -2}\phi ^{2}\left( \nabla u\right) ^{2}+
\end{equation*}%
\begin{equation}
+\func{div}\left( U_{5}+U_{6}\right) +\partial _{t}\left(
\dsum\limits_{i,j=1}^{n}a^{ij}\left( x\right) u_{i}u_{j}\phi ^{2}\right) ,
\label{4.30}
\end{equation}%
where $\func{div}U_{5}$ and $\func{div}U_{6}$ are given in (\ref{4.240}) and
(\ref{4.29}) respectively.

Recall that up to this point we have worked with $\mu \geq \mu _{0}.$ Now,
however, we set everywhere above and below $\mu =\mu _{0}.$ By (\ref{3.5}) $%
\varphi ^{-2\mu _{0}-2}\left( 1/4\right) ^{2\mu _{0}+2}\leq 1$ in $G.$
Multiplying both sides of (\ref{4.30}) by $\left( 1/4\right) ^{2\mu
_{0}+2}/\left( 2\lambda \mu _{0}\right) ,$ we obtain%
\begin{equation*}
\frac{1}{2\lambda \mu _{0}4^{2\mu _{0}+2}}\left( u_{t}-L_{0}u\right)
^{2}\phi ^{2}\geq \frac{u_{t}^{2}}{4^{2\mu _{0}+2}\left( \lambda \mu
_{0}\right) }\phi ^{2}+\frac{C}{4^{2\mu _{0}+2}\left( \lambda \mu
_{0}\right) }\dsum\limits_{i,j=1}^{n}u_{ij}^{2}\phi ^{2}-
\end{equation*}%
\begin{equation}
-\frac{C}{2}\lambda \mu _{0}\phi ^{2}\left( \nabla u\right) ^{2}+\func{div}%
\left( \frac{1}{4^{2\mu _{0}+2}\left( \lambda \mu _{0}\right) }\left(
U_{5}+U_{6}\right) \right) +  \label{4.31}
\end{equation}%
\begin{equation*}
+\partial _{t}\left( \frac{1}{4^{2\mu _{0}+2}\left( \lambda \mu _{0}\right) }%
\dsum\limits_{i,j=1}^{n}a^{ij}\left( x\right) u_{i}u_{j}\phi ^{2}\right) ,
\end{equation*}%
where $\func{div}U_{5}$ and $\func{div}U_{6}$ are given in (\ref{4.240}) and
(\ref{4.29}) respectively.

\subsection{The final estimate}

\label{sec:8.9}

Sum up (\ref{4.22}) with (\ref{4.31}) and then multiply both sides of the
resulting inequality by the number $d$ defined in (\ref{3.161}). We obtain
the target estimate (\ref{3.15}). Formulas (\ref{3.16}) and (\ref{3.160})
for $V_{t}$ and $\func{div}U$ follow from a combination of (\ref{4.7})-(\ref%
{4.60}), (\ref{4.13}), (\ref{4.16})-(\ref{4.22}), (\ref{4.240}) and (\ref%
{4.29})-(\ref{4.31}). $\square $


\begin{thebibliography}{99}
\bibitem{Alifanov} O.M. Alifanov, \emph{Inverse Heat Conduction Problems},
Springer, New York, 1994.

\bibitem{BukhKlib} A.L. Bukhgeim and M.V. Klibanov, Uniqueness in the large
of a class of multidimensional inverse problems, \emph{Soviet Mathematics
Doklady}, 17\textbf{, }244-247, 1981.

\bibitem{Das} B.B. Das, F. Liu and R. R. Alfano, Time-resolved fluorescence
and photon migration studies in biomedical and model random media, \emph{%
Reports on Progress in Physics}, 60, 227-292, 1997.

\bibitem{ImYam1} O.Y. Imanuvilov and M. Yamamoto, Lipschitz stability in
inverse parabolic problems by the Carleman estimate, \emph{Inverse Problems}%
, 14, 1229-1245 (1998).

\bibitem{ImYam2} O.Y. Imanuvilov and M. Yamamoto, Inverse parabolic problems
by Carleman estimates with data taken initial or final time moment of
observation, \emph{Inverse Problems and Imaging, published online, }doi:
10.3934/ipi.2023036, 2023.

\bibitem{Isakov} V. Isakov, \emph{Inverse Problems for Partial Differential
Equations}, Springer, New York, 2006.

\bibitem{Klib84} M.V. Klibanov, Inverse problems in the `large' and Carleman
bounds. Differential Equations, 20, 755-760, 1984.

\bibitem{Klib92} M.V. Klibanov, Inverse problems and Carleman estimates, 
\emph{Inverse Problems}, 8, 575--596, 1992.

\bibitem{Ksurvey} M.V. Klibanov, Carleman estimates for global uniqueness,
stability and numerical methods for coefficient inverse problems, \emph{J.
Inverse and Ill-Posed Problems} , 21, 477-510, 2013.

\bibitem{KY} M.V. Klibanov and A.G. Yagola, Convergent numerical methods for
parabolic equations with reversed time via a new Carleman estimate, \emph{%
Inverse Problems,} 35, 115012, 2019.

\bibitem{KLZ} M.V. Klibanov, J. Li and W. Zhang, Convexification for an
inverse parabolic problem, \emph{Inverse Problems,} 36, 085008, 2020.

\bibitem{KL} M.V. Klibanov and J. Li, \emph{Inverse Problems and Carleman
Estimates: Global Uniqueness, Global Convergence and Experimental Data}, De
Gruyter, 2021\emph{.}

\bibitem{Ladpar} O.A. Ladyzhenskaya, V.A. Solonnikov and N.N. Uraltceva, 
\emph{Linear and Quasilinear Equations of Parabolic Type}, AMS, Providence,
RI, 1968.

\bibitem{Lad} O.A. Ladyzhenskaya, \emph{Boundary Value Problems of
Mathematical Physics}, Springer, 1985.

\bibitem{LRS} M.M. Lavrentiev, V.G. Romanov and S.P. Shishatskii, \emph{%
Ill-Posed Problems of Mathematical Physics and Analysis}, AMS, Providence,
RI, 1986.

\bibitem{Lay} R. Y. Lai and Q. Li, Parameter reconstruction for general
transport equation, \emph{SIAM J. Math. Analasys}, 52, 2734-2758, 2020.

\bibitem{Nov1} R.G. Novikov, The inverse scattering problem on a fixed
energy level for the two-dimensional Schr\"{o}dinger operator\emph{,} \emph{%
J. Functional Analysis}, 103\textbf{,} 409-463 1992.

\bibitem{Nov2} R. G. Novikov, $\partial -$bar approach to approximate
inverse scattering at fixed energy in three dimensions, \emph{International
Math. Research Peports}, 6,\textbf{\ }287-349 2005.

\bibitem{Prilepko} A. I. Prilepko, D.G. Orlovsky and I.A. Vasin, \emph{%
Methods for Solving Inverse Problems in Mathematical Physics}, Marcel
Dekker, Inc., New York 1999.

\bibitem{Rom} V.G. Romanov, \emph{Inverse Problems of Mathematical Physics},
VNU Press, Utrecht, 1986.
\end{thebibliography}
\end{document}